\documentclass[useAMS,usenatbib]{mnras}
\usepackage[T1]{fontenc}
\usepackage{ae,aecompl}
\usepackage{pdflscape}
\usepackage{graphicx}
\usepackage{longtable}
\usepackage{lscape}
\usepackage{amssymb}
\usepackage{endnotes} 
\usepackage{footnote}
\usepackage{subfigure}
\usepackage{amsmath}
\usepackage{txfonts}
\usepackage{natbib}
\usepackage{url}
\usepackage{breakurl}
\usepackage{helvet}
\usepackage[flushleft]{threeparttable}
\def\aj{{AJ}}%Astronomical Journal
\def\araa{{ARA\&A}}% Annual Review of Astron and Astrophys
\def\apj{{ApJ}}% Astrophysical Journal
\def\apjl{{ApJ}}% Astrophysical Journal, Letters
\def\apjs{{ApJS}}% Astrophysical Journal, Supplement
\def\aap{{A\&A}}% Astronomy and Astrophysics
\def\mnras{{MNRAS}}% Monthly Notices of the RAS
\def\HI{H{\small{I}}~}
\def\HII{H{\small{II}}~}

\title[Wolf-Rayet galaxies]{Narrowband H$\alpha$ imaging of nearby Wolf-Rayet galaxies}
\author[A. Paswan et al.]{A.~Paswan$^{1}$ \thanks{E-mail:
paswanabhishek@iucaa.in}, Kanak Saha$^{1}$ \thanks{E-mail:kanak@iucaa.in} and A.~Omar$^{2}$ \thanks{E-mail:aomar@aries.res.in}\\ $^{1}$Inter-University Centre for Astronomy and Astrophysics, Ganeshkhind, Post Bag 4, Pune 411007, India \\$^{2}$Aryabhatta Research Institute of Observational Sciences, Manora Peak, Nainital 263002, India\\ }

\date{Accepted ------------, Received ------------; in original form ------------}
\pagerange{\pageref{firstpage}--\pageref{lastpage}} \pubyear{}
      
\begin{document}
\label{firstpage}
%\date{Accepted ------------, Received ------------; in original form ------------}
\pagerange{\pageref{firstpage}--\pageref{lastpage}} \pubyear{}
\maketitle

\begin{abstract}

We present narrowband H$\alpha$ imaging of nearby Wolf-Rayet (WR) galaxies known as a subset of starburst galaxies. The H$\alpha$ images have been used to show morphology of star-forming regions in galaxies, which leads to speculate that the studied galaxies have most likely experienced merger or interaction with low luminous dwarf galaxies or \HI clouds. We further derive the H$\alpha$ based star formation rates (SFRs) in galaxies using our H$\alpha$ observations. These SFRs are well-correlated with SFRs derived using other indicators at far-ultraviolet, far-infrared and 1.4-GHz radio wavebands. It is noticed that the infrared excess (IRX) method gives the best SFR estimates, consistent with different models predication. These models also predict that the sample galaxies have probably gone through a continuous star formation at least for 1 Gyr over which the recent ($\textless$ 10 Myr) star formation has taken place in WR phase. This study presents Main-Sequence (MS) relation for nearby WR galaxies for the first time. This derived MS relation is found to be similar to previously known MS relation for normal nearby star-forming galaxies, suggesting that WR systems evolve in a similar fashion as normal star-forming galaxies evolve.

\end{abstract}


%%%%%%%%%%%%%%%%%%%%%%%%%%%%%%%%%%%%%%%%%%%%%%%%%%%%%%%%%%%%%%%%%%%%%%%%%%%%%%%%%%%%%%%%%%%%%%%%%%%%%%%%%%%%%%%%%%%%%%%%%%%%%%%%%%%%%%%%%%%%%%%%%%%%%%

\section{Introduction}

A subset of starburst \HII galaxies known as Wolf-Rayet (WR) galaxies are classified based on broad optical emission features of He{\small{II}} $\lambda$4686 and C{\small{IV}} $\lambda$5808 lines, originated in the stellar winds from a substantial population of WR stars \citep{1976MNRAS.177...91A,1982ApJ...261...64O,1991ApJ...377..115C}. Such features in WR galaxies are often seen after 2 to 5 Myr of initial star formation for a short duration (t$_{WR}$ $\leq$ 0.5 Myr), until WR population end their lives in supernovae \citep{2005A&A...429..581M}. Due to a large population of WR stars, a very high ratio of WR to O-type stars in WR galaxies suggests that star formation in these systems takes place in burst mode in a short period with a rather flat initial mass function (IMF). This leads to classify the WR galaxies as starburst systems \citep{1995A&A...303..440C}. All these properties of WR galaxies reveals that they host ongoing star formation activities ($\textless$ 10 Myr), and are ideal systems to study onset of star formation and its triggering mechanism.

In order to measure star formation rate (SFR), there are several tracers such as far-ultraviolet (FUV), H$\alpha$ line emission, far-infrared (FIR) and radio continuum etc. However, these indicators trace different star formation histories in galaxies and predict different estimates of SFR due to their different sensitivity over ongoing physical processes in the interstellar medium (ISM) of galaxies. The details of these tracers are widely discussed in the literature \citep[e.g.,][]{1992ARA&A..30..575C,1998ARA&A..36..189K,2007ApJ...666..870C,2007ApJS..173..267S,2011ApJ...737...67M}. Apart from the SFR estimation, the knowledge of its triggering mechanism is equally important. Although there are various mechanisms which trigger star formation in galaxies, one of these mechanisms is tidal interaction/merger of galaxies or \HI clouds. This mechanism can be easily traced through studying morphological features of star-forming regions and stellar component in galaxies \citep[e.g.,][]{2010A&A...521A..63L,2012MNRAS.419.1051L,2014A&A...566A..71L,2015ApJ...815L..17M}.     
 
Despite the WR galaxies are interesting systems for probing physical processes involved in star formation in galaxies, very few detailed studies of WR galaxies have been found in the literature \citep{2010A&A...521A..63L,2014MNRAS.439..157K,2016MNRAS.462...92J}. It is therefore worth to expend such studies using a larger population of WR galaxies. We here report our H$\alpha$ and r-band imaging of a sample of new 13 nearby WR galaxies, selected from the catalogues of \citet{1999A&A...341..399S} and \citet{2008A&A...485..657B}. The selected galaxies have diverse morphological types varying from low-mass dwarf irregular galaxies to large spiral galaxies. They are chosen such that their redshifted H$\alpha$ emission could be observed with the narrow-band H$\alpha$ filter available on the observing telescope. Furthermore, the galaxies were constrained to a declination $\textgreater$ - 25$^{0}$, so that they could be observed for sufficient integration time in a single observing night. The galaxies are brighter than 17.5 mag in Sloan Digital Sky Survey (SDSS) r-band and have sub-solar metallicities. Their stellar masses lie in the range of $10^{5.5} - 10^{10.5}$ M$_{\odot}$.

In this study, we present the morphologies of ionized gas and stellar component in galaxies to see any imprint of tidal interaction/merger features using H$\alpha$ and r-band observations. We exploit our observed H$\alpha$ data in the estimation of global H$\alpha$-based SFRs which are further compared with other SFRs from different indicators at FUV, FIR and radio wavebands. This study makes an effort to better understand the ongoing physical processes in WR galaxies by investigating the consistency between different SFR indicators. Here, we include more other sample of 45 WR galaxies from \citet[][; hereafter LS10]{2010A&A...521A..63L} and \citet[][; hereafter JO16]{2016MNRAS.462...92J}, to make our results statistically strong. The relevant cosmological parameter used through out this work is $H_{0}$ = 75 km s$^{-1}$.
 
\section{Observations and data reduction}

The selected galaxies were observed using the 1.3-m Devasthal Fast Optical Telescope (DFOT) operated by Aryabhatta Research Institute of Observational Sciences (ARIES), Nainital, India. The galaxies were imaged in broad r-band and narrow-band H$\alpha$ filters using a thinned, back-illuminated E2V 2k$\times$2k CCD (plate scale = 0.54$\arcsec$ pixel$^{-1}$) having pixels size of 13.5 $\mu$m. The central wavelength ($\lambda_{0}$) of the H$\alpha$ filter is at 6563 \AA~with a full width at half-maxima (FWHM) of $\sim$100 \AA. The observations were carried out on dark nights under the photometric sky conditions. At least one spectrophotometric standard star from \citet{1990AJ.....99.1621O} was observed in each night for calibration. The observed average H$\alpha$ sensitivity was found as $\sim$ 5 $\times$ 10$^{-15}$ erg~s$^{-1}$~cm$^{-2}$~arcsec$^{-2}$. The observation log is provided in Table~\ref{tab:01}. 

\begin{table*}
\centering
%\tiny
\caption{Summary of narrow (H$\alpha$) and broad (r) bands observations of the sample galaxies using 1.3-m DFOT.}
\vspace {0.3cm}
\begin{tabular}{cccccccc} \hline 
Galaxy name &	Date & Exposure in H$\alpha$ & Exposure in r & Seeing & Distance & $E_f(B-V)$ & $E_g(B-V)$ \\\hline
	       & [dd-mm-yyyy]  &	[sec] & [sec] & [arcsec] & [Mpc] & [mag] & [mag] \\ 
\hline
NGC 1140   & 27-01-2014 & 7200 & 1200 & 2.4 & 20.0 & 0.03 & 0.01\\   
NGC 2481   & 04-04-2014 & 5700 & 1600 & 2.6 & 28.8 & 0.06 & ---\\
Mrk 94     & 27-01-2014 & 9500 & 1260 & 2.6 & 09.8 & 0.03 & 0.07\\
UGC 5249   & 26-01-2014 &11100 & 1800 & 2.2 & 25.0 & 0.05 & 0.07\\
UGC 6526   & 13-04-2013 & 5400 & 1200 & 2.5 & 24.5 & 0.02 & 0.20\\
NGC 3755   & 27-01-2014 & 3700 & 1360 & 2.3 & 20.9 & 0.02 & 0.28\\
Mrk 750    & 30-04-2014 & 6300 & 1800 & 2.5 & 10.0 & 0.03 & 0.00\\
UGC 6805   & 28-04-2014 & 7500 & 2400 & 2.6 & 15.1 & 0.02 & 0.26\\
IC 745     & 04-04-2014 & 6600 & 1500 & 2.6 & 15.3 & 0.02 & 0.22\\
NGC 4496A  & 13-04-2013 & 1800 & 1200 & 2.5 & 23.1 & 0.02 & ---\\
NGC 4904   & 12-04-2013 & 4200 &  540 & 2.4 & 15.7 & 0.02 & 0.28\\
NGC 5147   & 12-04-2013 & 3900 &  900 & 2.3 & 14.5 & 0.02 & 0.16\\
Izw 97	   & 12-04-2013 & 1800 &  420 & 2.3 & 33.6 & 0.02 & 0.24\\      
\hline 
\end{tabular} \\
\begin{flushleft}
{\footnotesize {{\bf Note:} The value of $E_f(B-V)$ is taken from NED based on \citet{2011ApJ...737..103S} recalibration of \citet{1998ApJ...500..525S} extinction map. The value $E_{g}(B-V)$ is estimated from the line flux ratio of $f_{H\alpha}$/$f_{H\beta}$ using the relation given by \citet{1989S&T....78..491O}. The value of $f_{H\alpha}$/$f_{H\beta}$ is taken from MPA-JHU (Max Planck Institute for Astrophysics and Johns Hopkins University) emission line database.}}
\end{flushleft}
\label{tab:01}
\end{table*}

The observed data were reduced using the standard tasks under the {\small{IRAF}} (Image Reduction and Analysis Facility) developed by the National Optical Astronomy Observatory (NOAO). The reduction process includes flat-fielding, cosmic-ray removal, alignment of the frames, continuum subtraction from the H$\alpha$ band image along with removal of other lines contamination in both broad and narrow-band filters. Finally, the H$\alpha$ photometry was done. Apart from our own observations, the ancillary data at FUV, FIR and 1.4 GHz radio continuum wavebands were taken from the Galaxy Evolution Explore (GALEX), Infrared Astronomical Satellite (IRAS and Faint Images of Radio Sky at Twenty-centimeters (FIRST)/NRAO VLA Sky Survey (NVSS), respectively. The resulting photometric fluxes in optical and UV wavebands were corrected for both Galactic and internal extinction using reddening values provided in Table~\ref{tab:01}, assuming \citet{1989ApJ...345..245C} extinction curve of $R_{V}$ = 3.1. All the steps involved in data reduction and analysis were performed exactly in the same manner as described in JO16.

\section{Results} \label{Sect-3}

\begin{figure*}
\centering
\includegraphics[width=4.0cm,height=4.0cm]{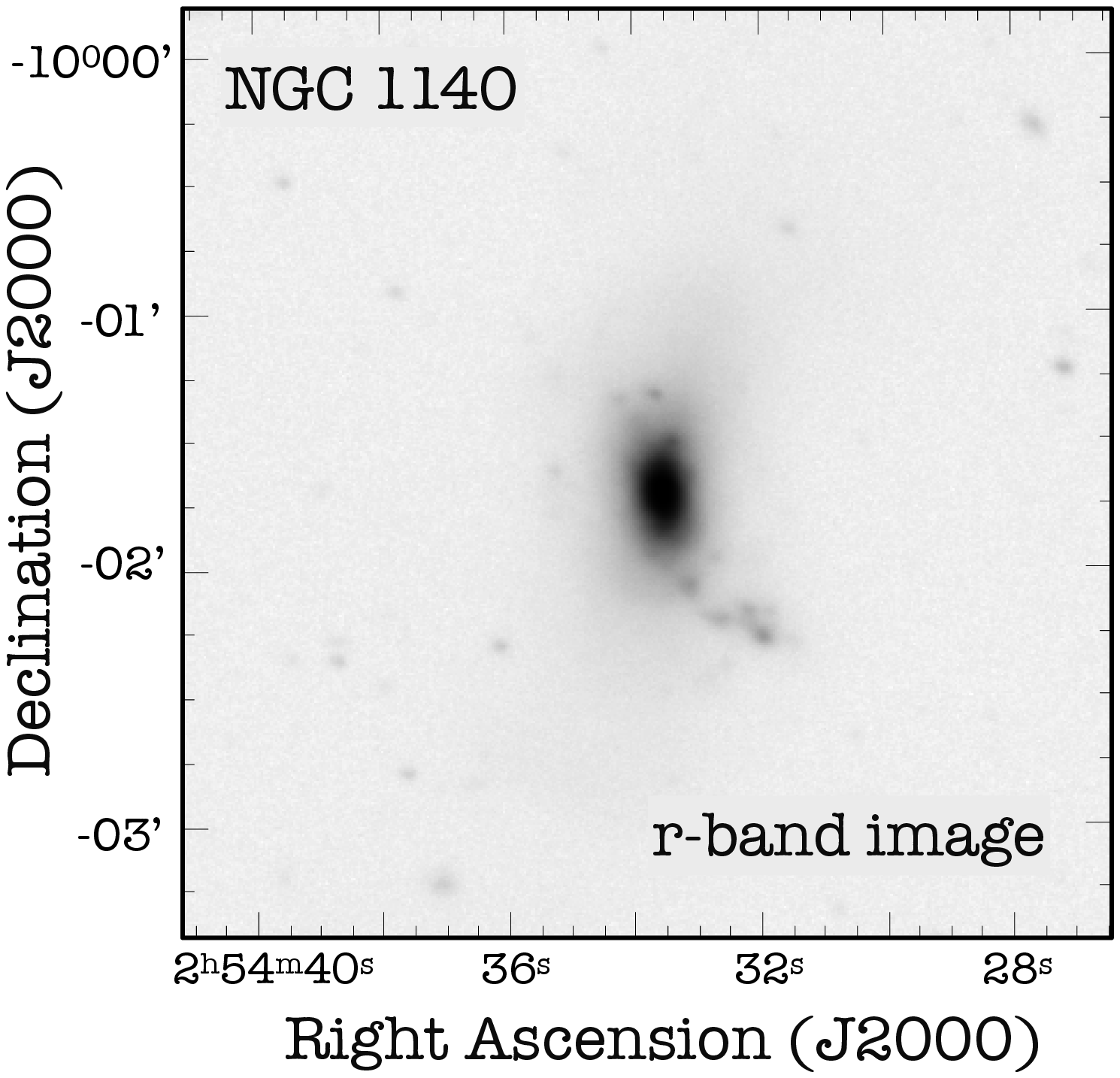}
\includegraphics[width=4.0cm,height=4.0cm]{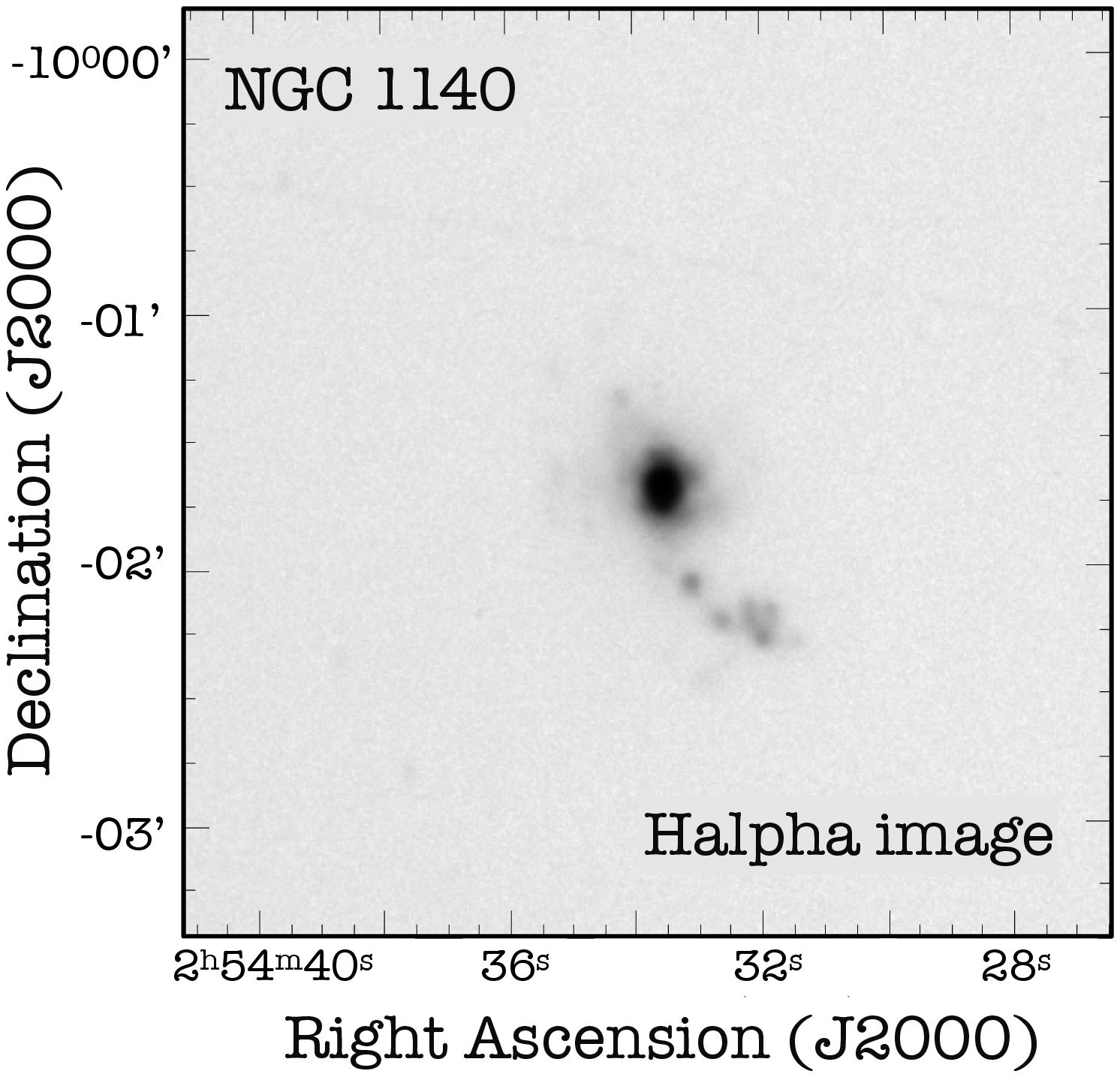}
\includegraphics[width=4.0cm,height=4.0cm]{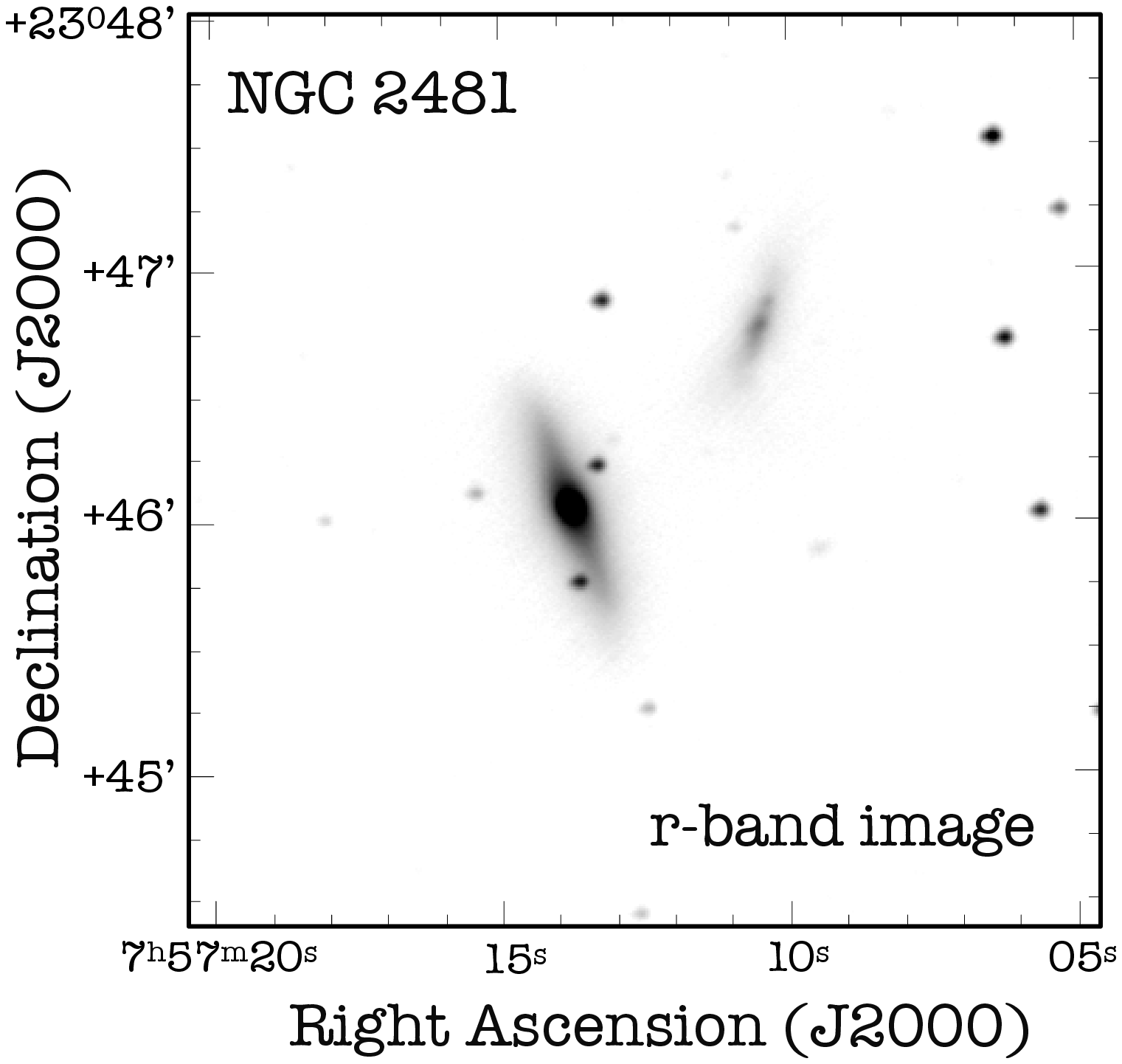}
\includegraphics[width=4.0cm,height=4.0cm]{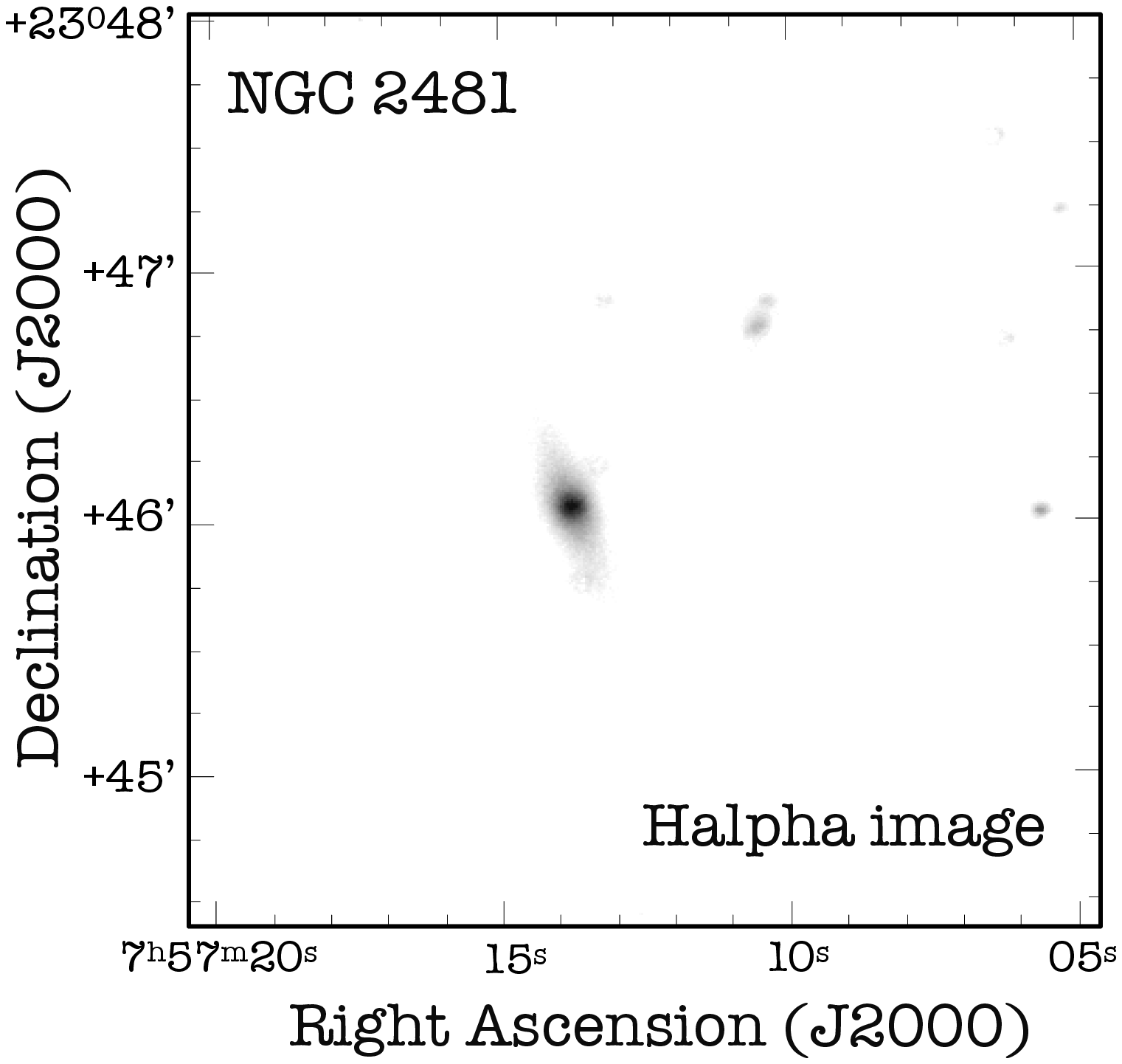}
\includegraphics[width=4.0cm,height=4.0cm]{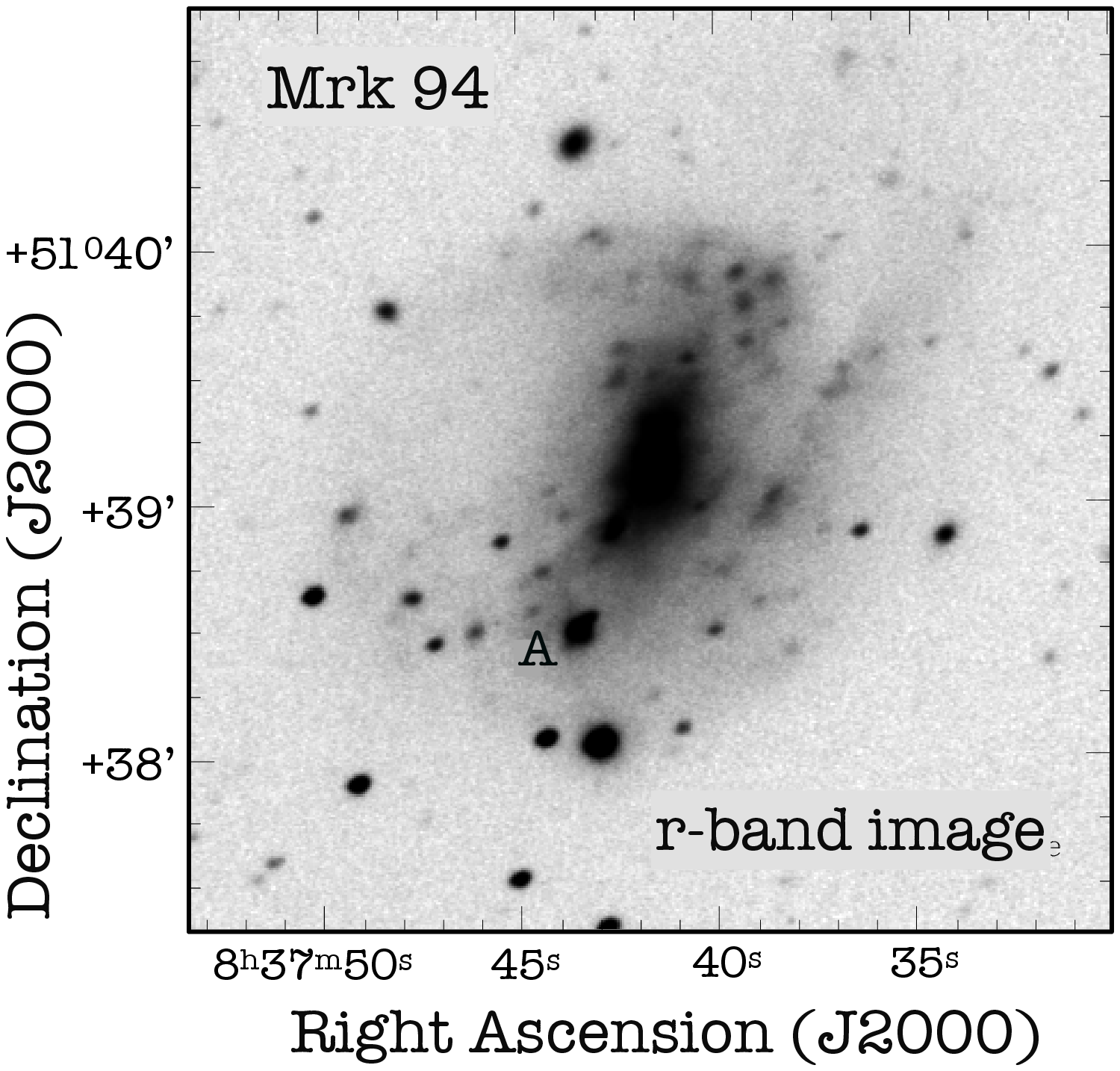}
\includegraphics[width=4.0cm,height=4.0cm]{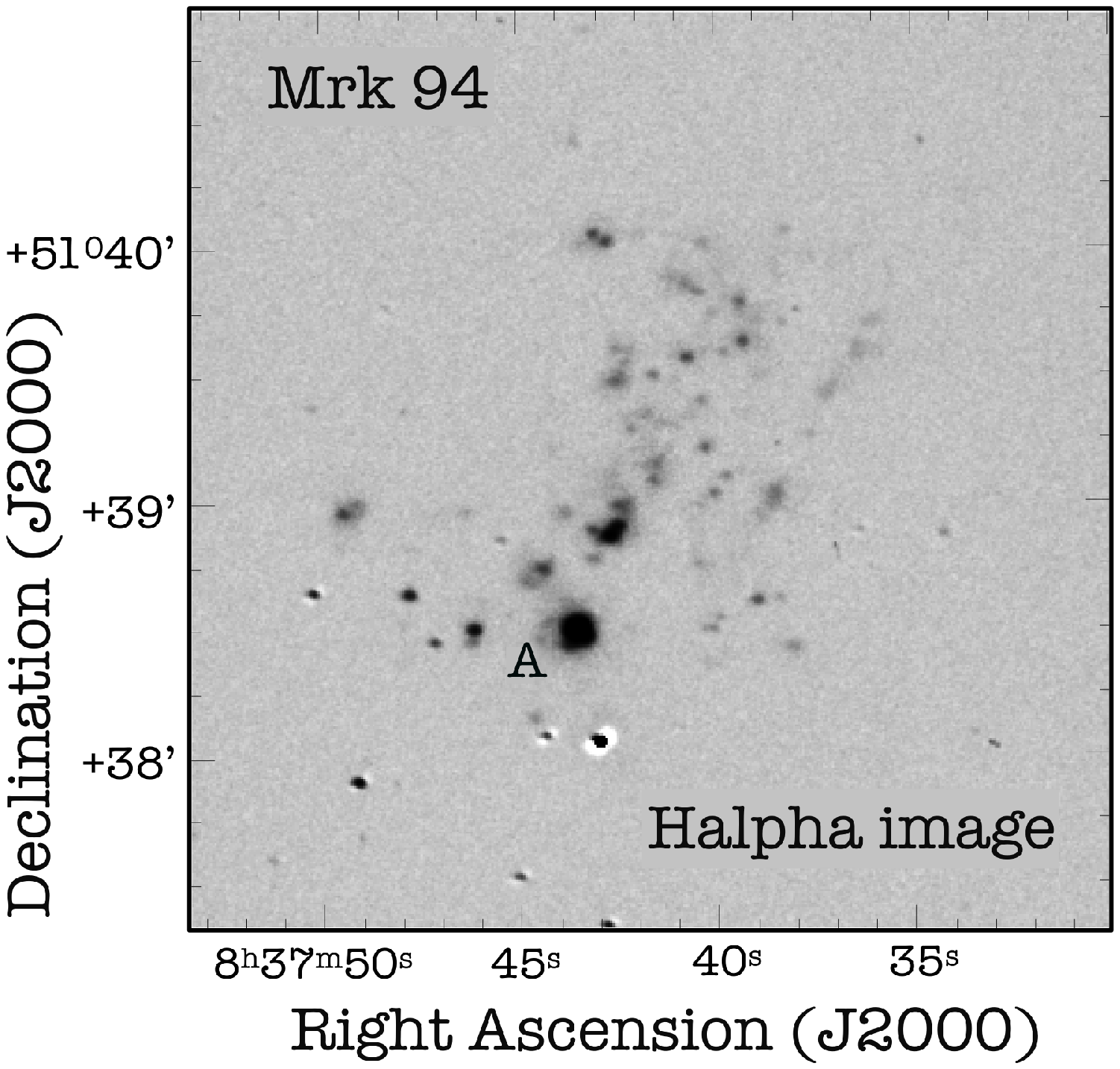}
\includegraphics[width=4.0cm,height=4.0cm]{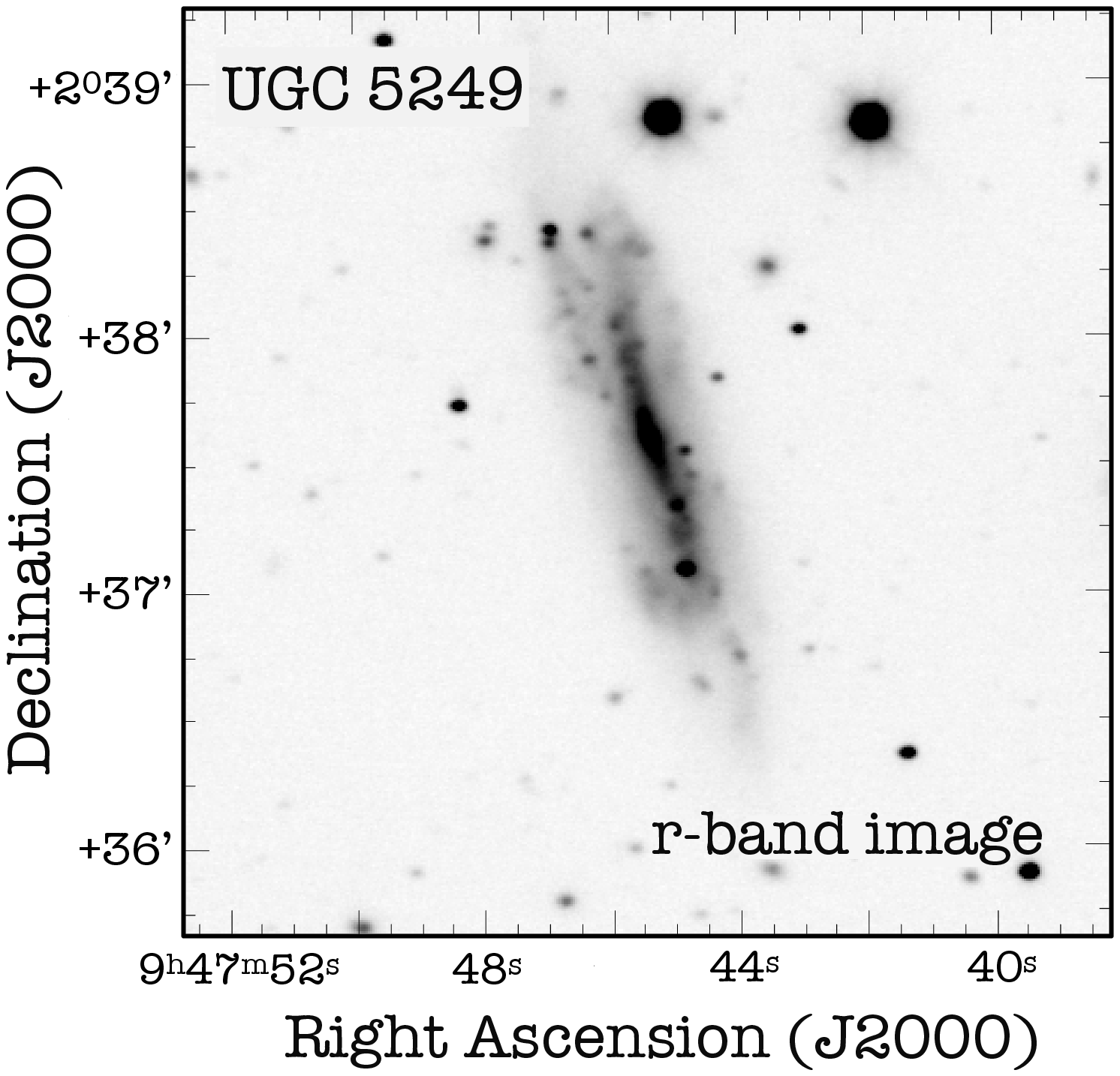}
\includegraphics[width=4.0cm,height=4.0cm]{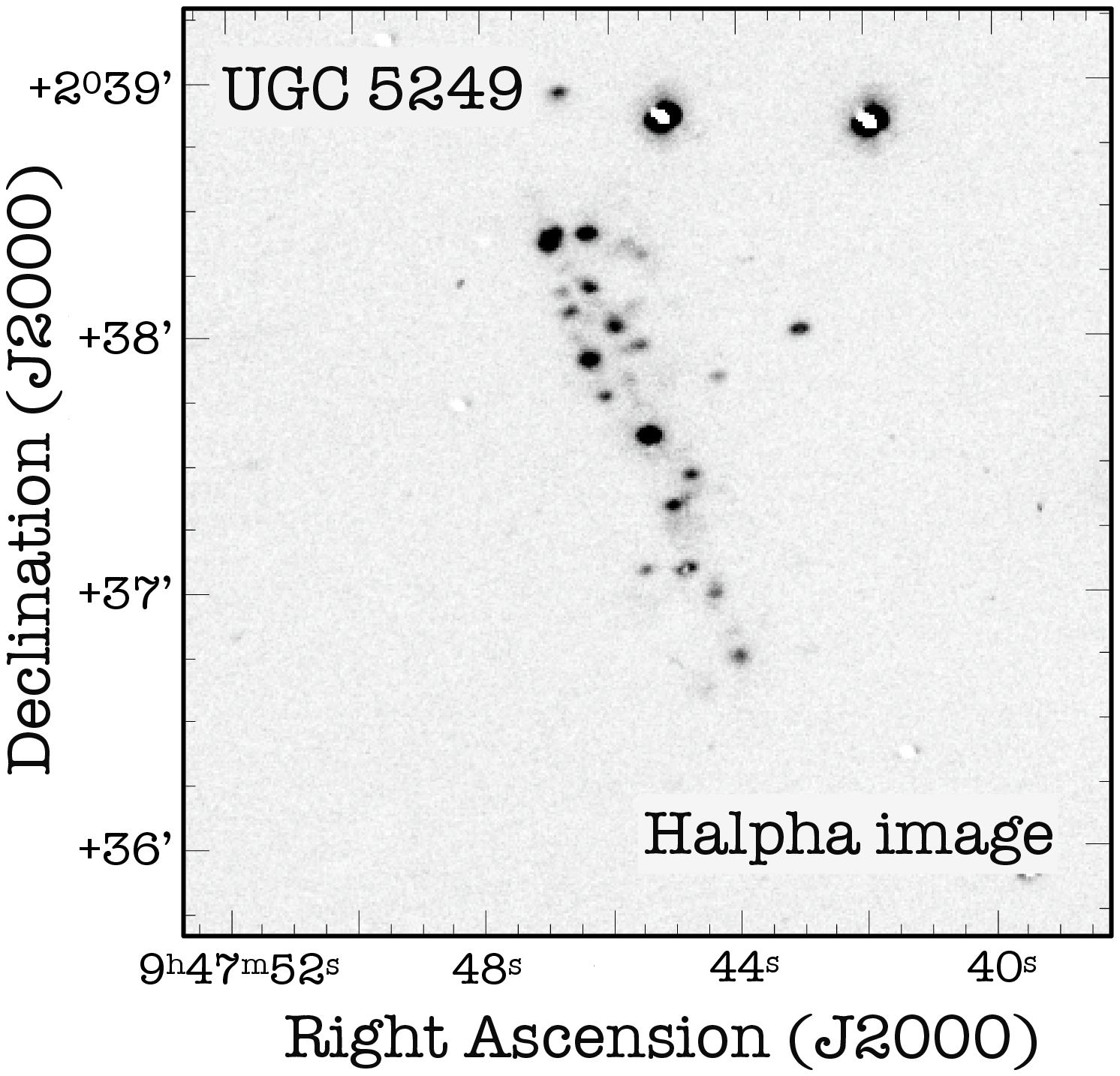}
\includegraphics[width=4.0cm,height=4.0cm]{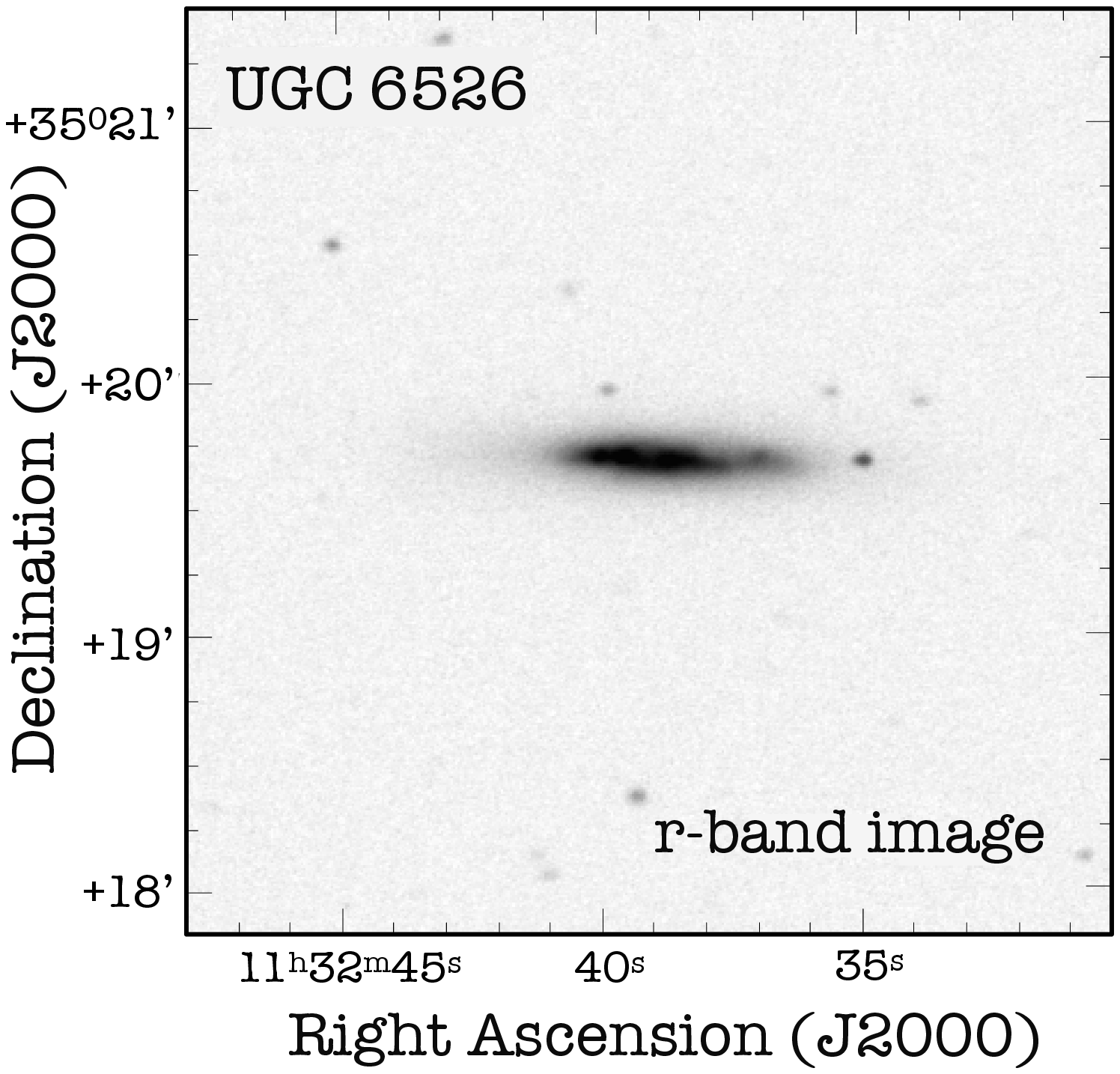}
\includegraphics[width=4.0cm,height=4.0cm]{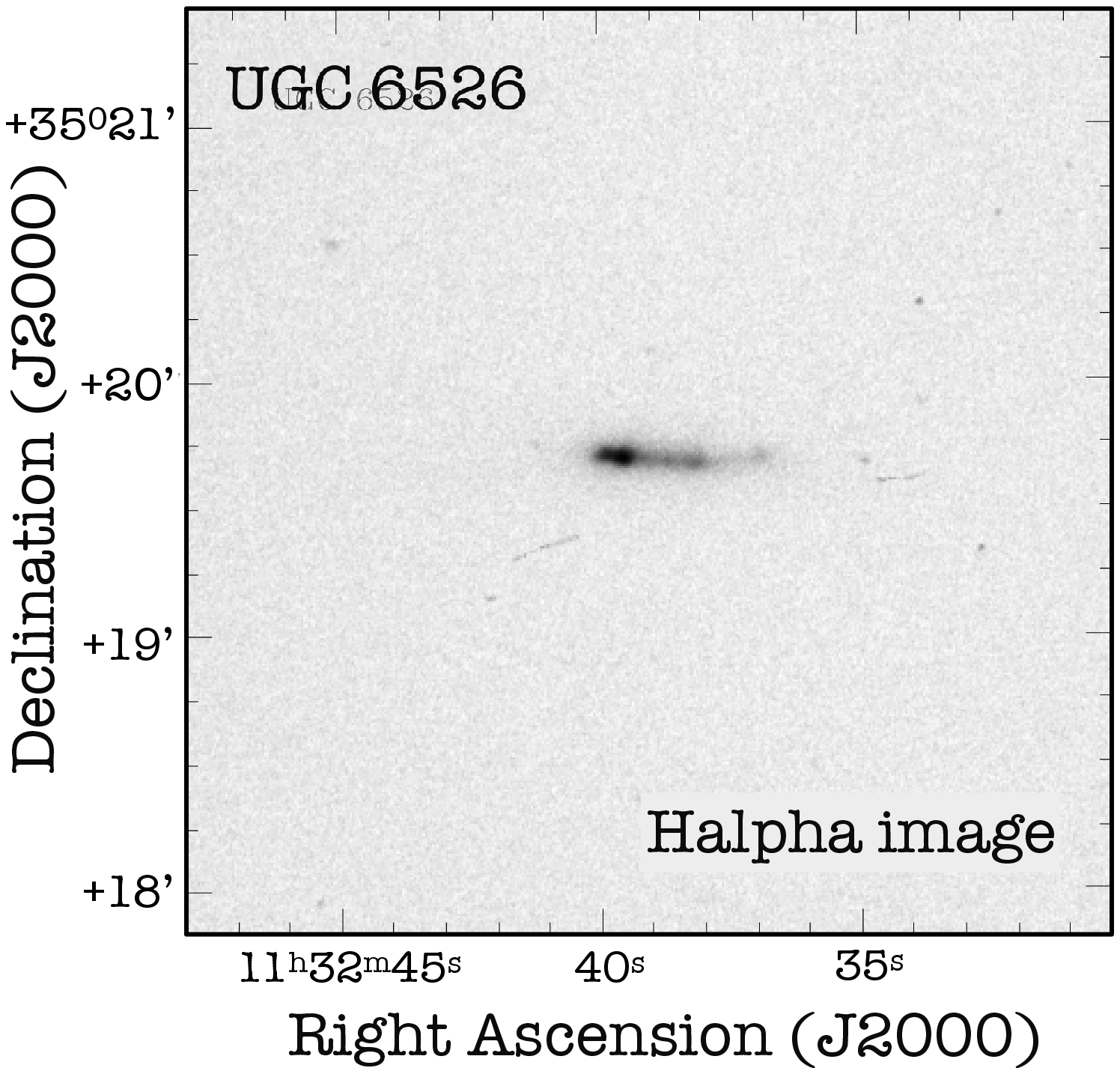}
\includegraphics[width=4.0cm,height=4.0cm]{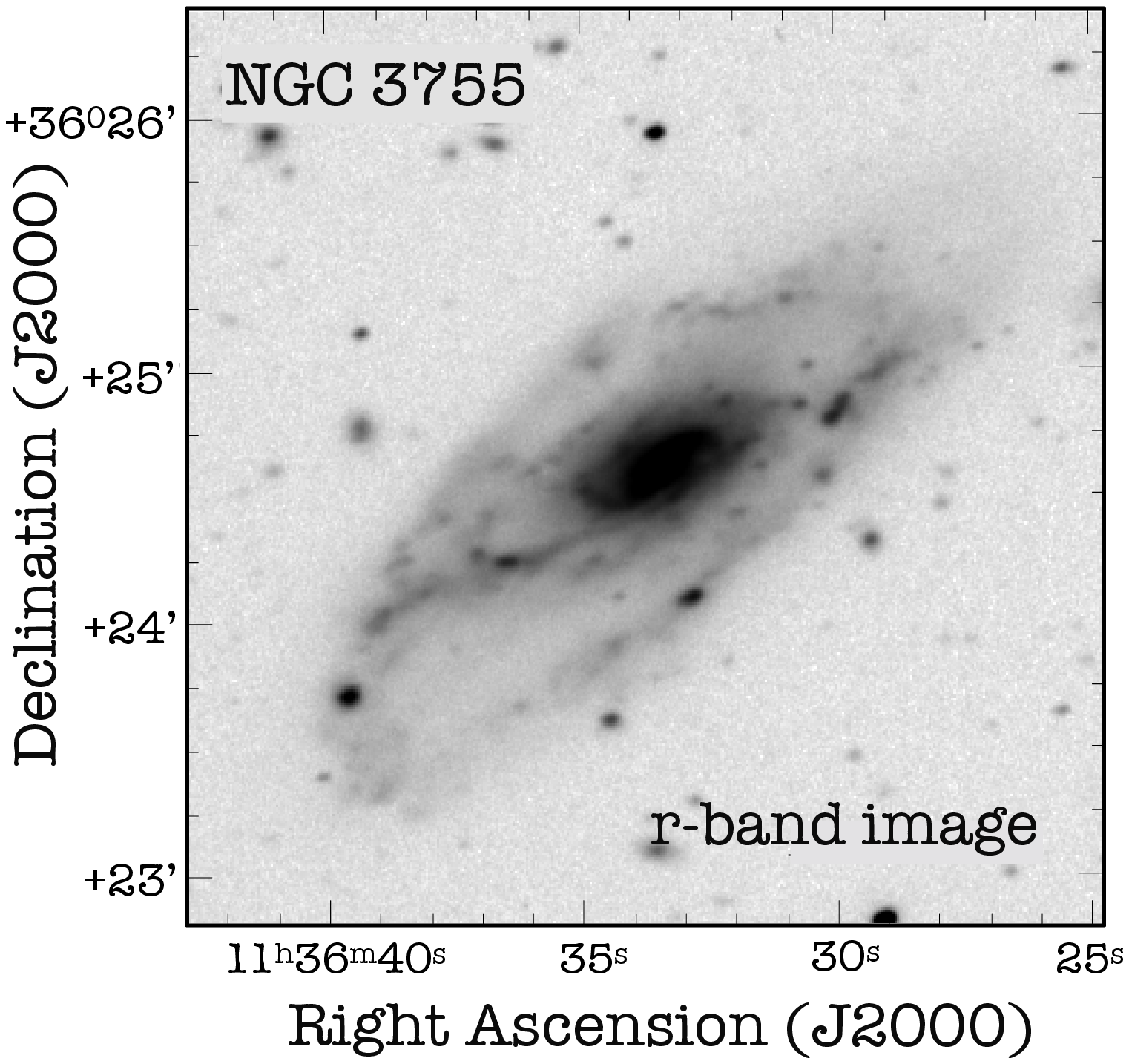}
\includegraphics[width=4.0cm,height=4.0cm]{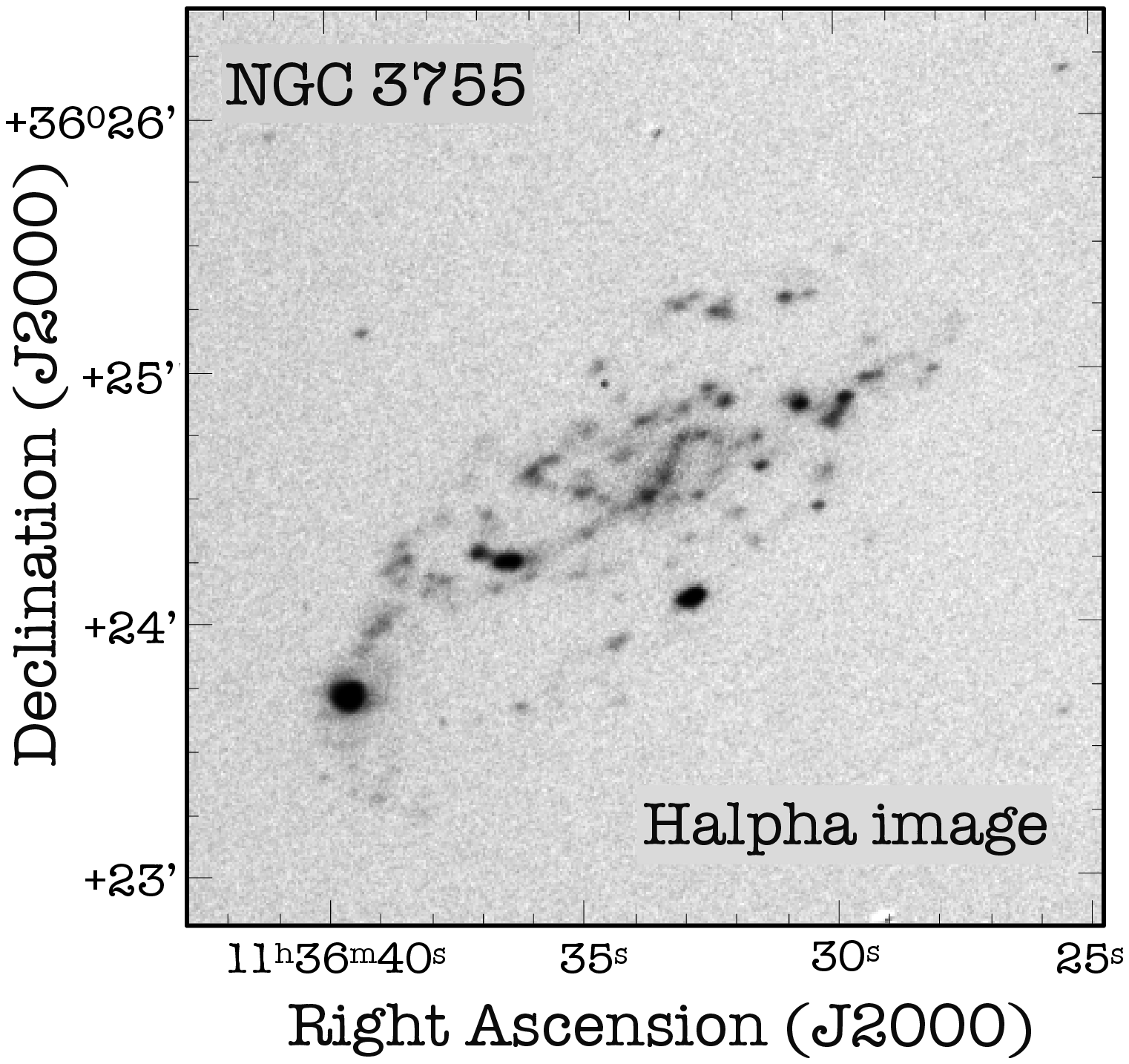}
\includegraphics[width=4.0cm,height=4.0cm]{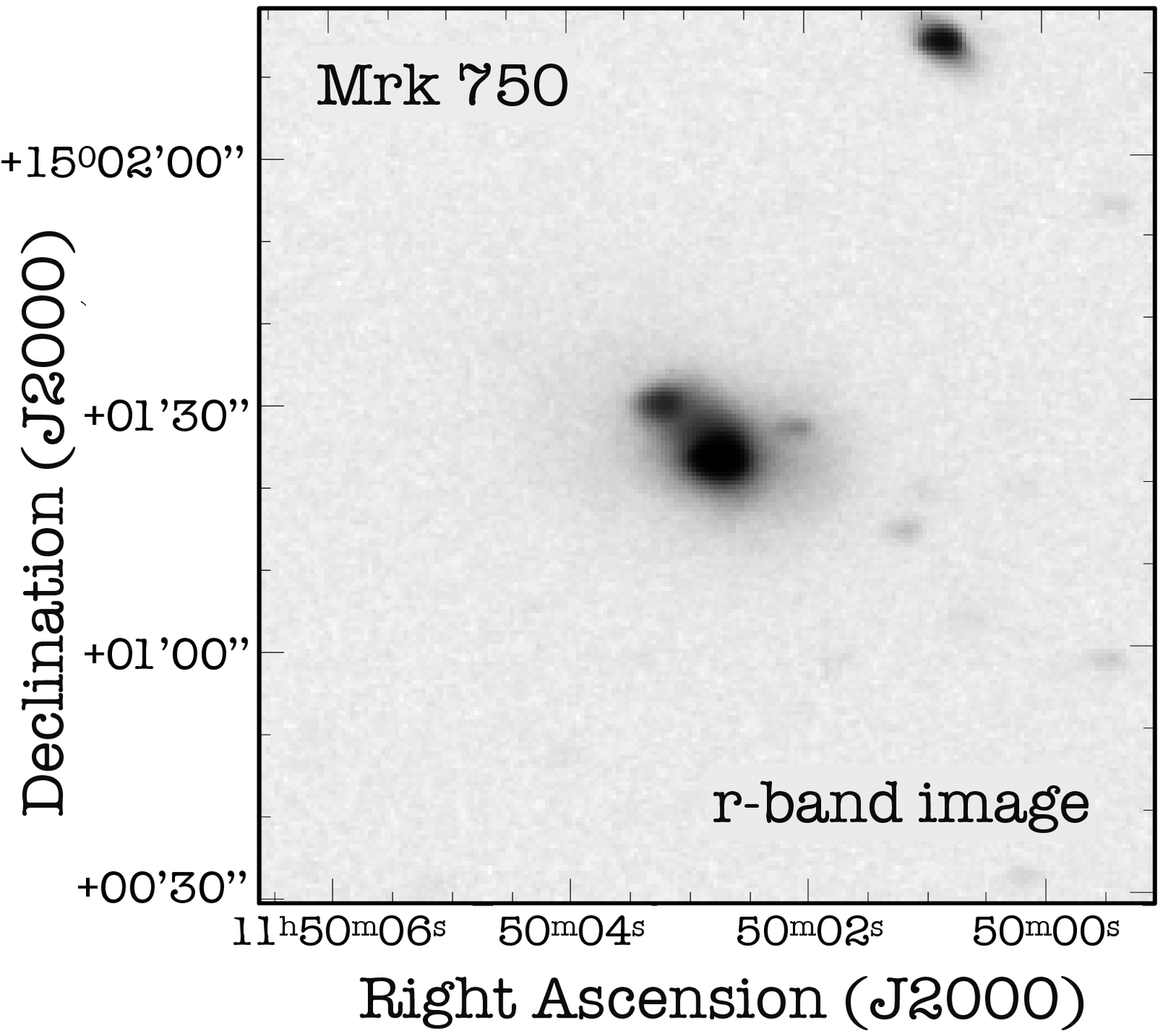}
\includegraphics[width=4.0cm,height=4.0cm]{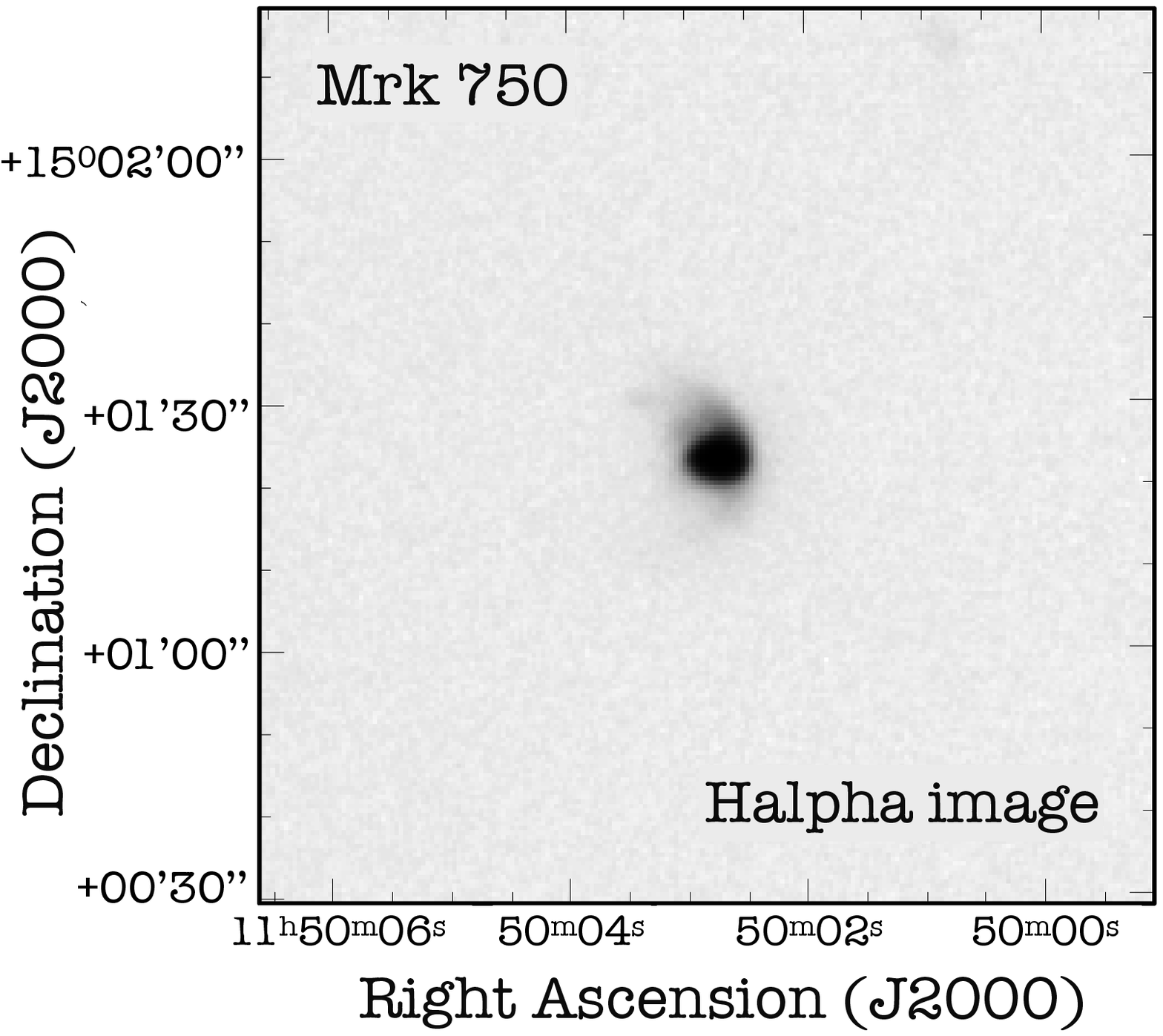}
\includegraphics[width=4.0cm,height=4.0cm]{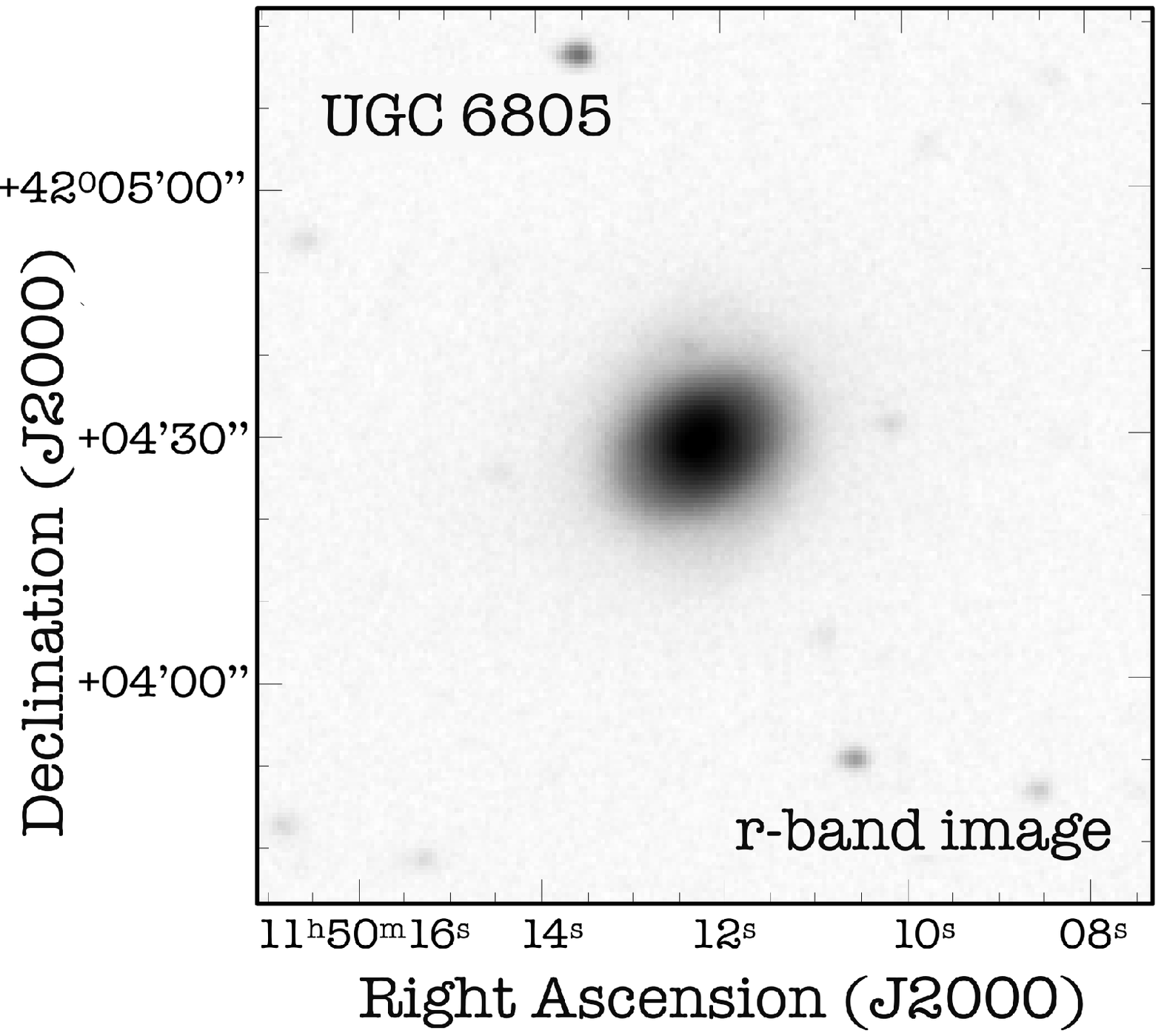}
\includegraphics[width=4.0cm,height=4.0cm]{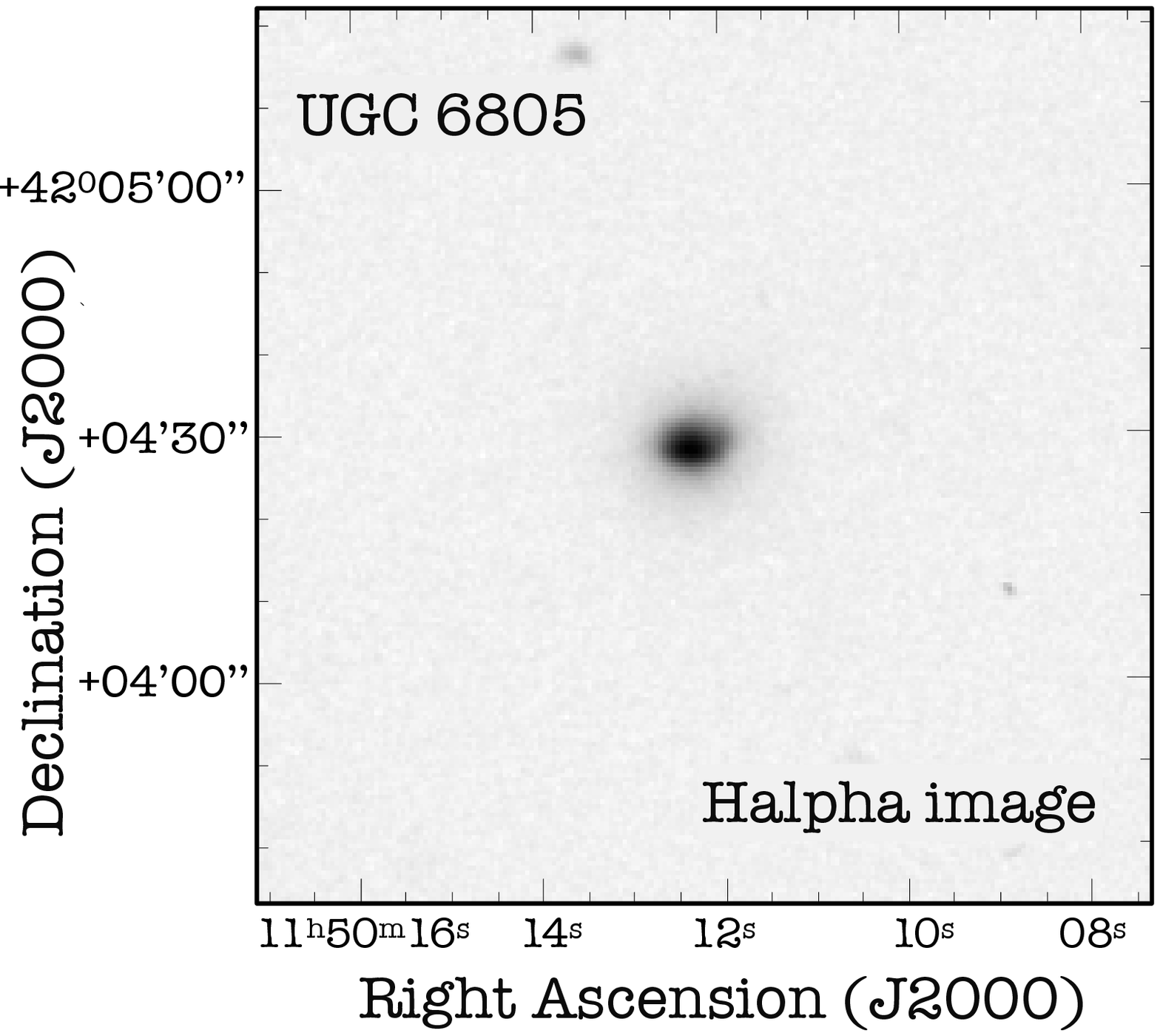}
\includegraphics[width=4.0cm,height=4.0cm]{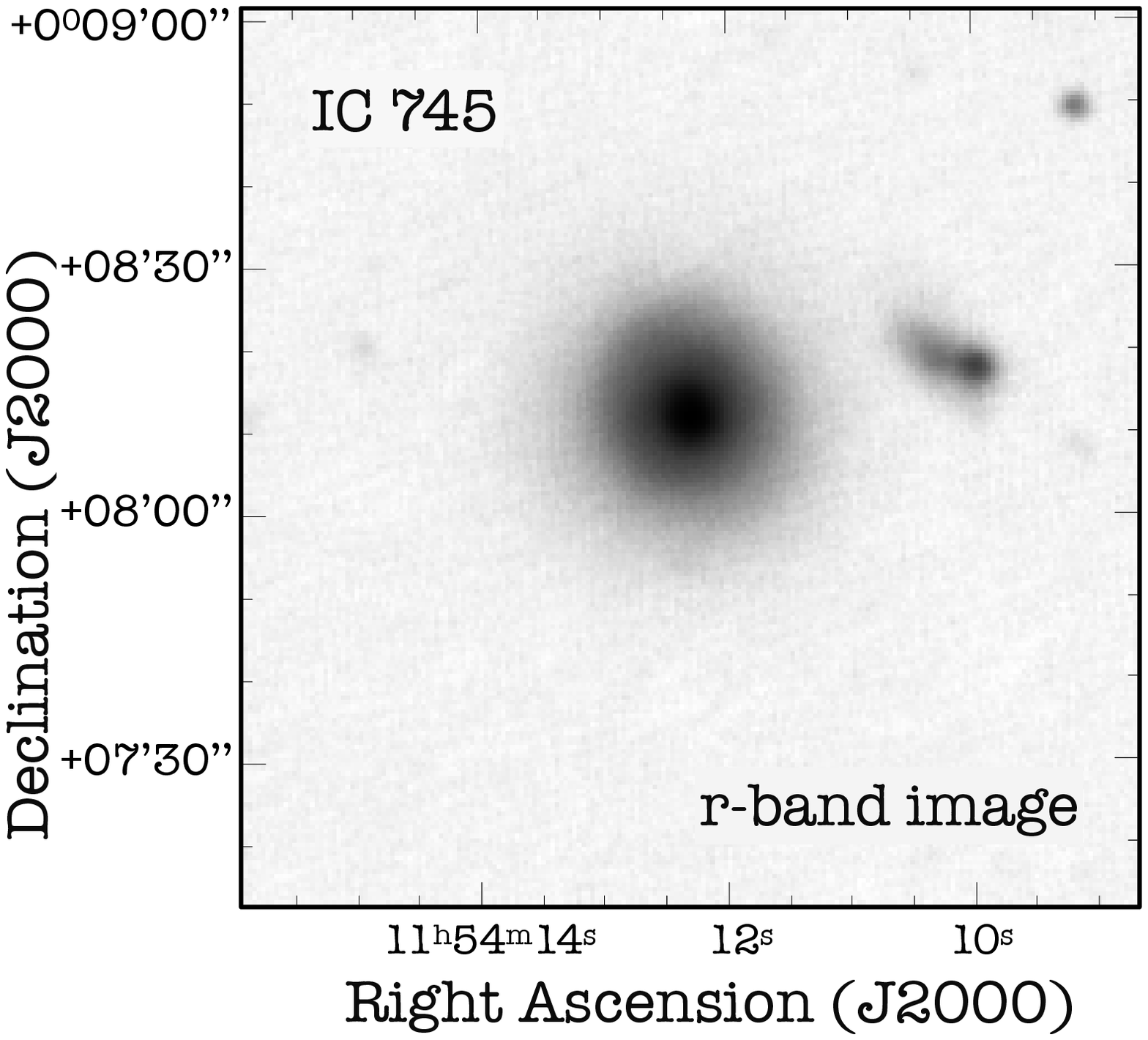}
\includegraphics[width=4.0cm,height=4.0cm]{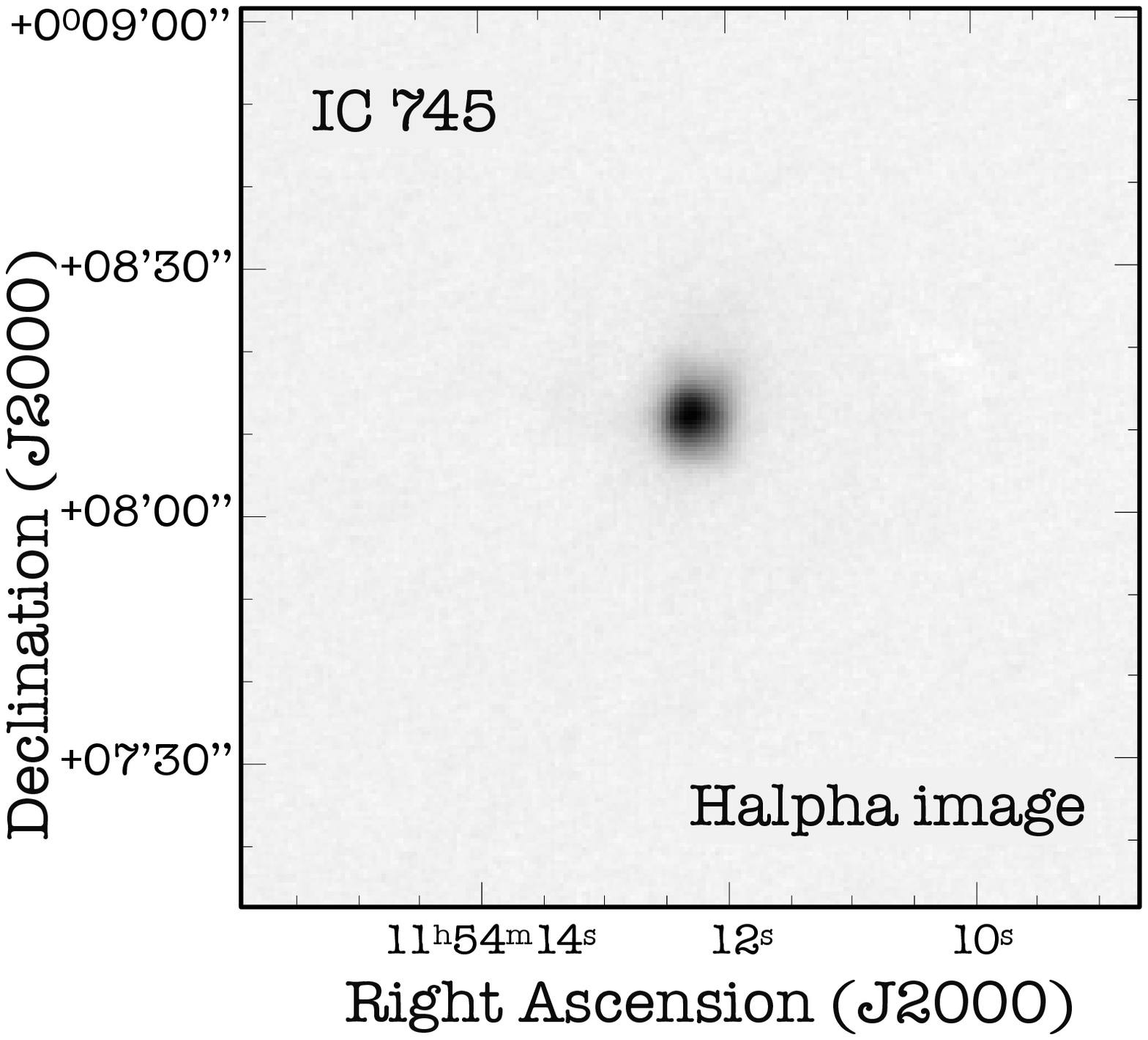}
\includegraphics[width=4.0cm,height=4.0cm]{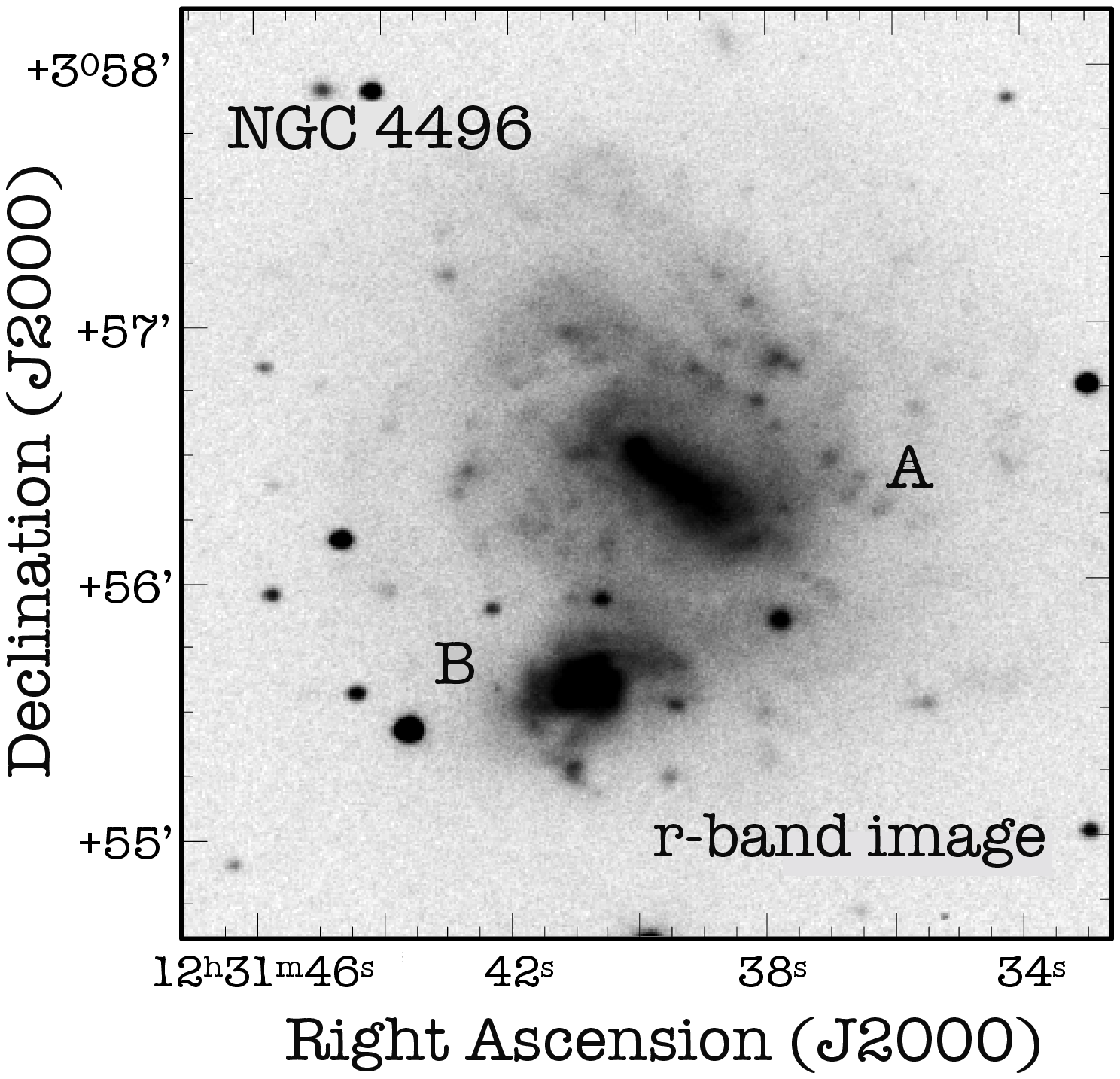}
\includegraphics[width=4.0cm,height=4.0cm]{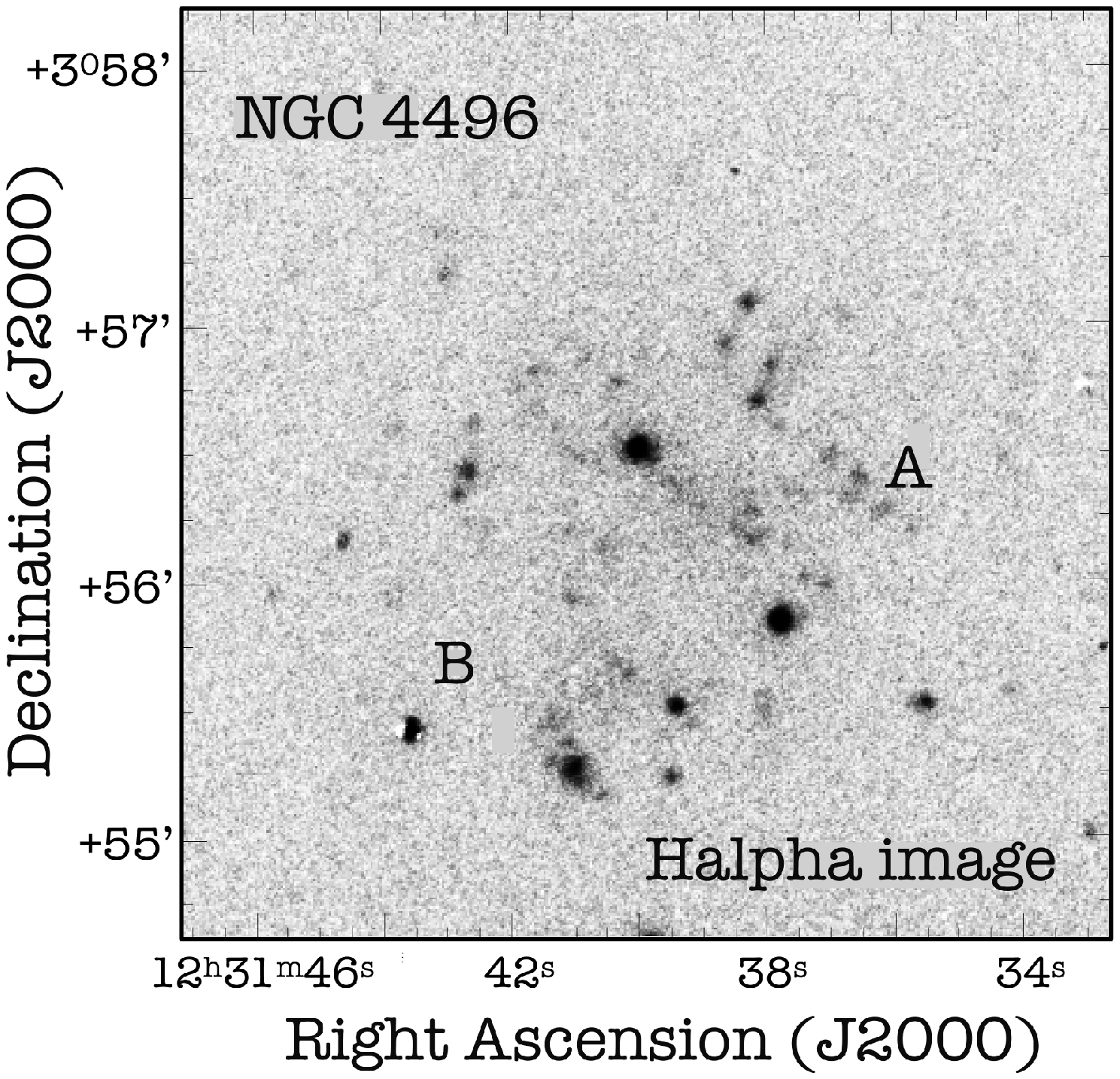}
\includegraphics[width=4.0cm,height=4.0cm]{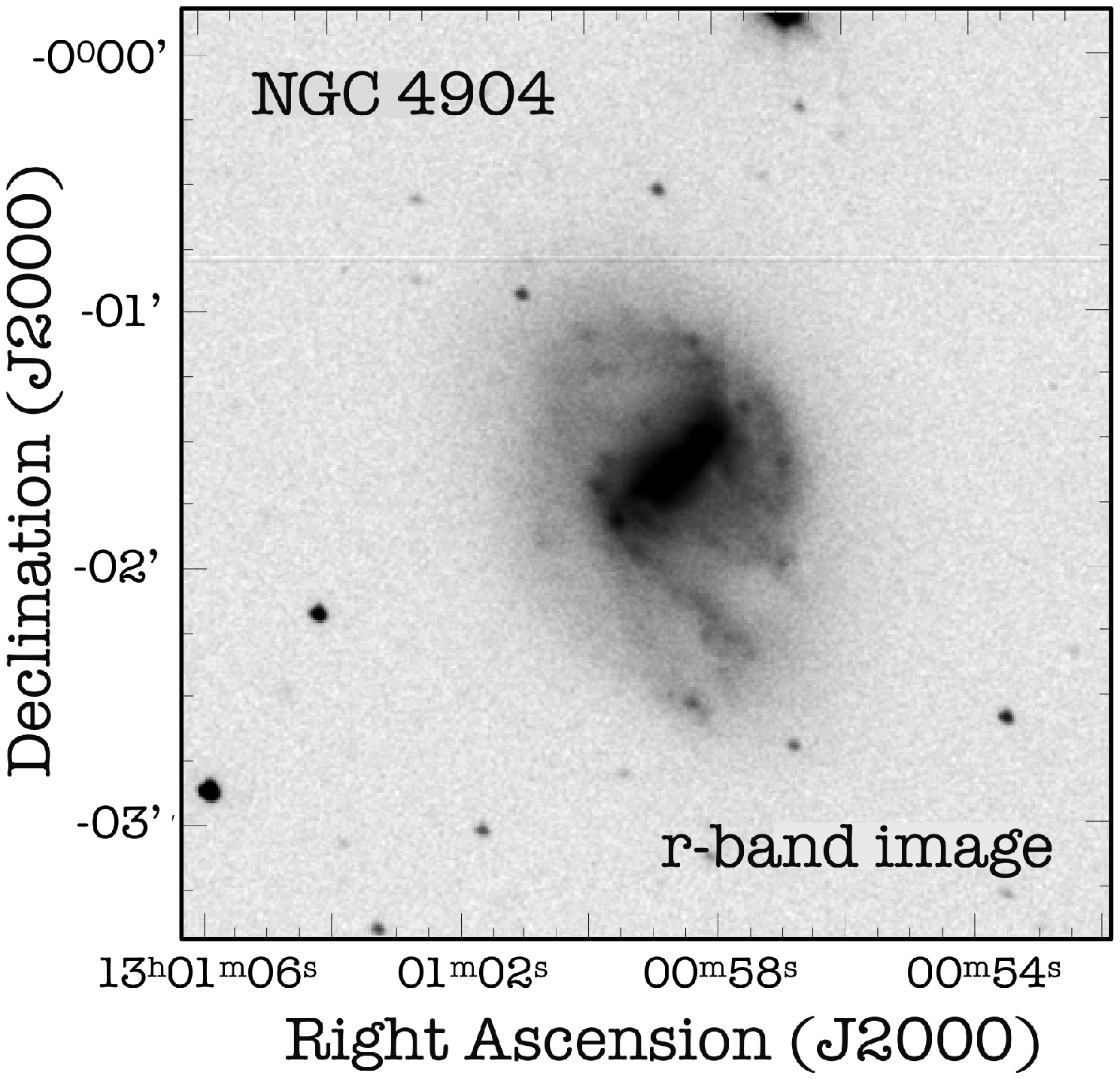}
\includegraphics[width=4.0cm,height=4.0cm]{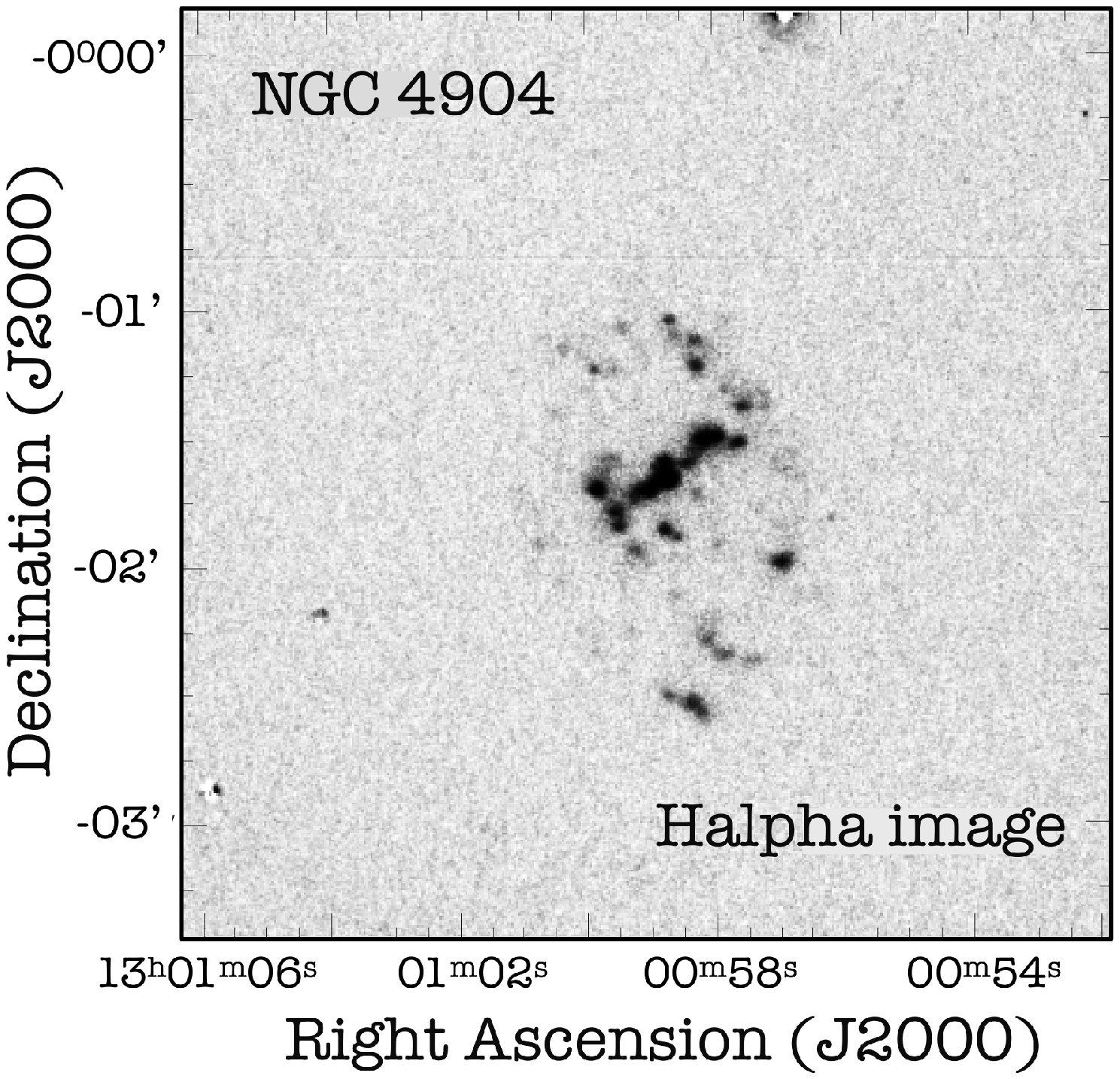}
\includegraphics[width=4.0cm,height=4.0cm]{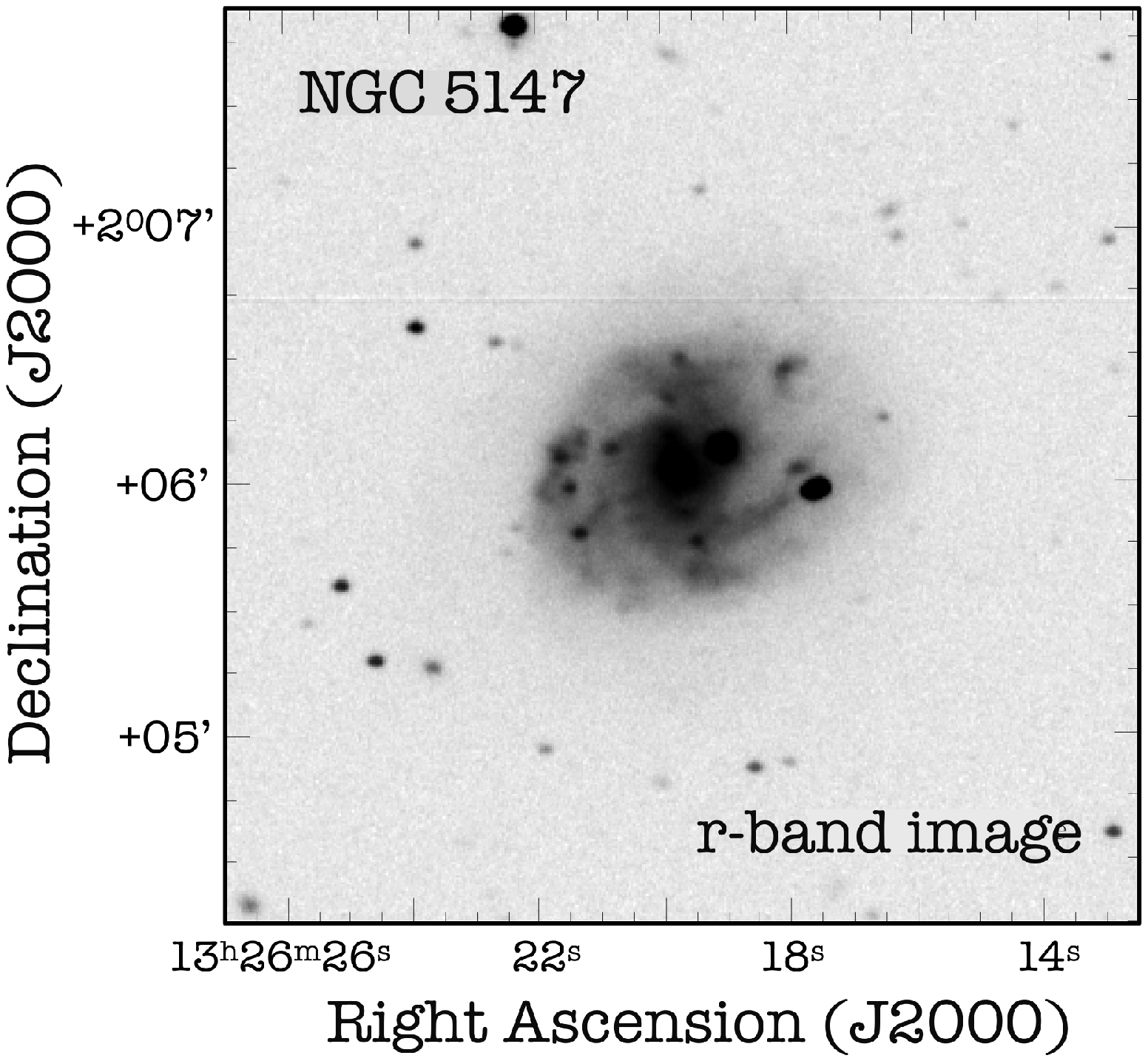}
\includegraphics[width=4.0cm,height=4.0cm]{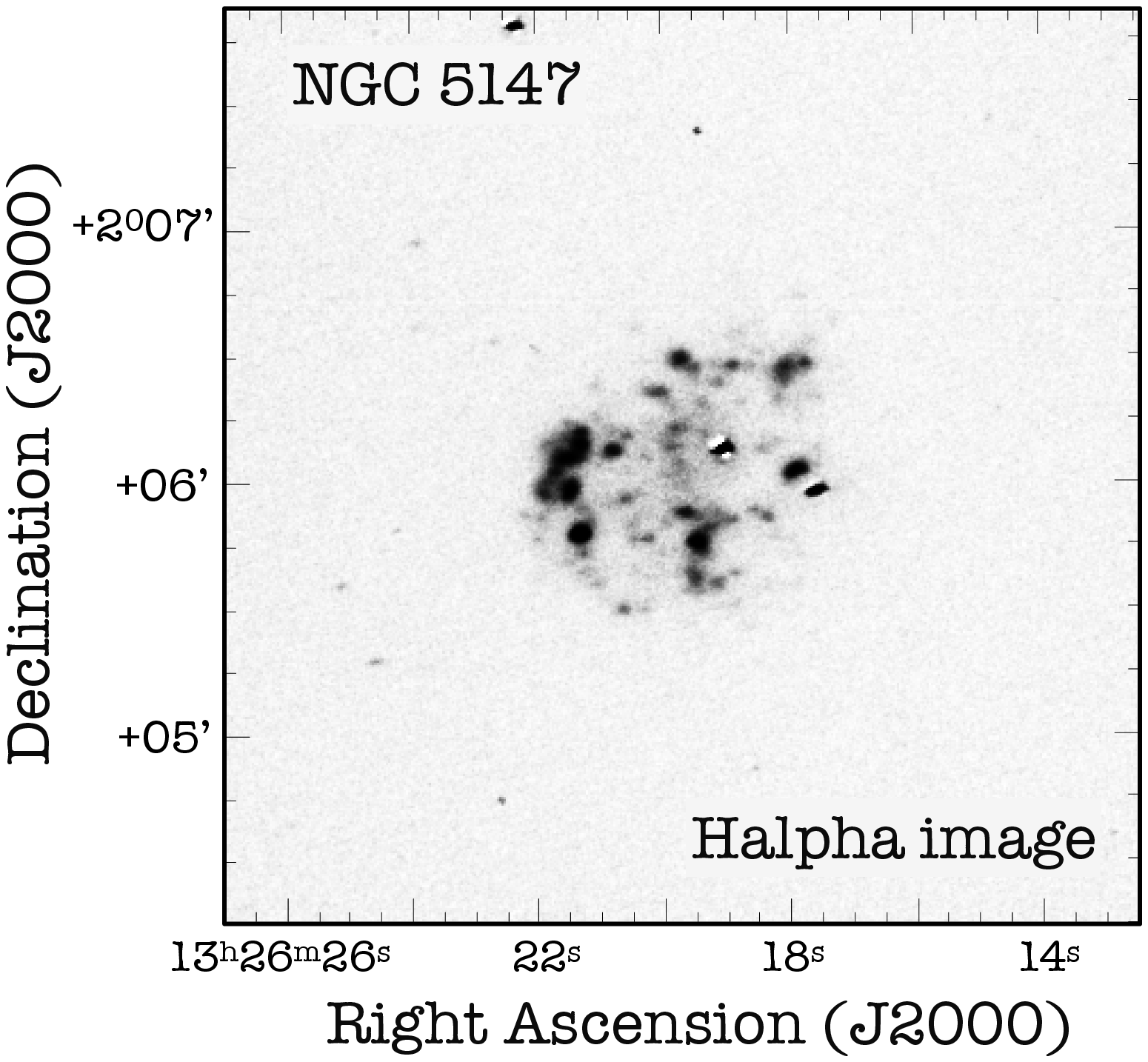}
\caption{The broad $r$-band (left) and continuum subtracted H$\alpha$ (right) images of WR galaxies in our sample.}
\label{fig:01}
\end{figure*}
\addtocounter{figure}{-1}
\begin{figure*}
\centering
\includegraphics[width=4.0cm,height=4.0cm]{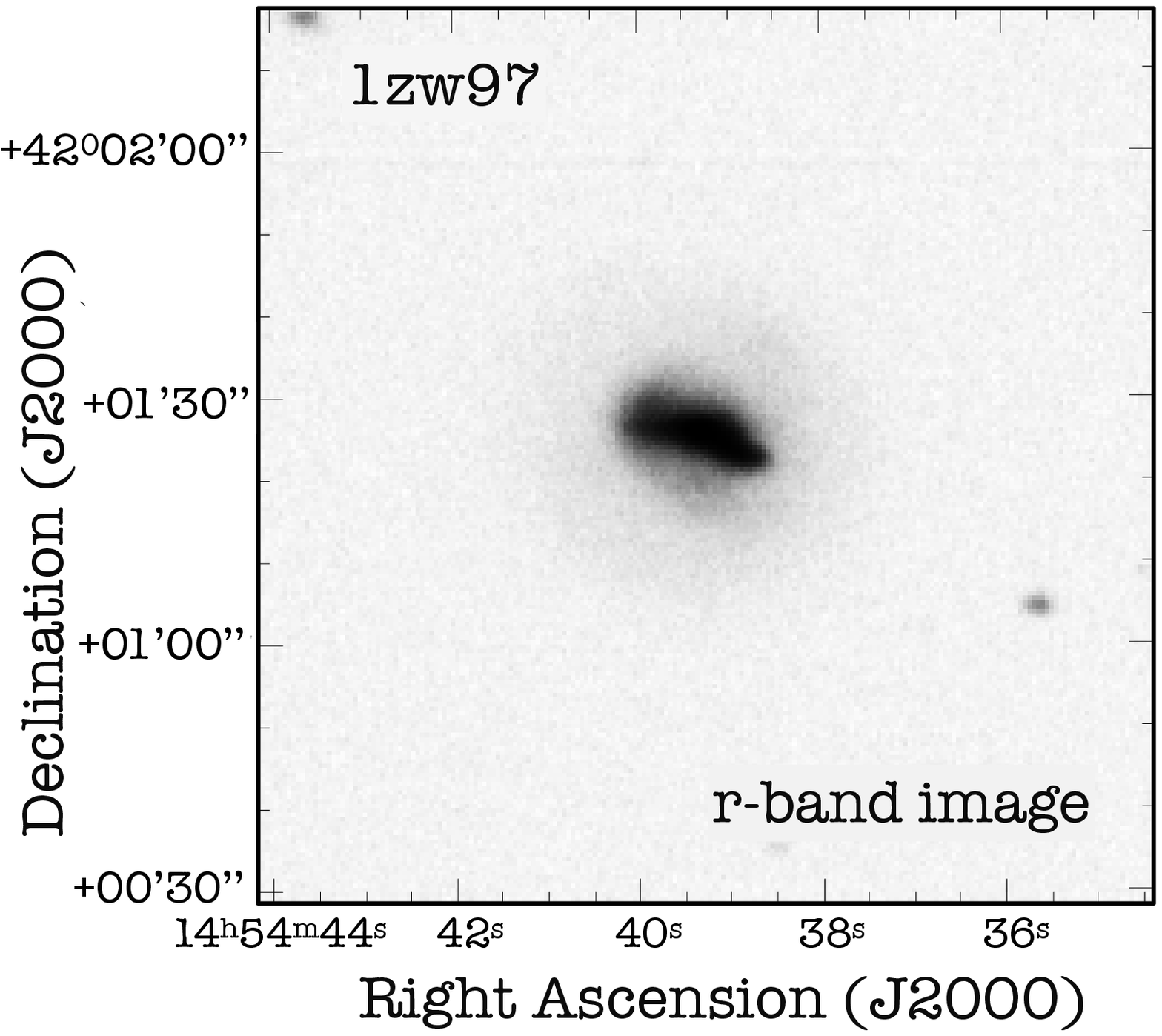}
\includegraphics[width=4.0cm,height=4.0cm]{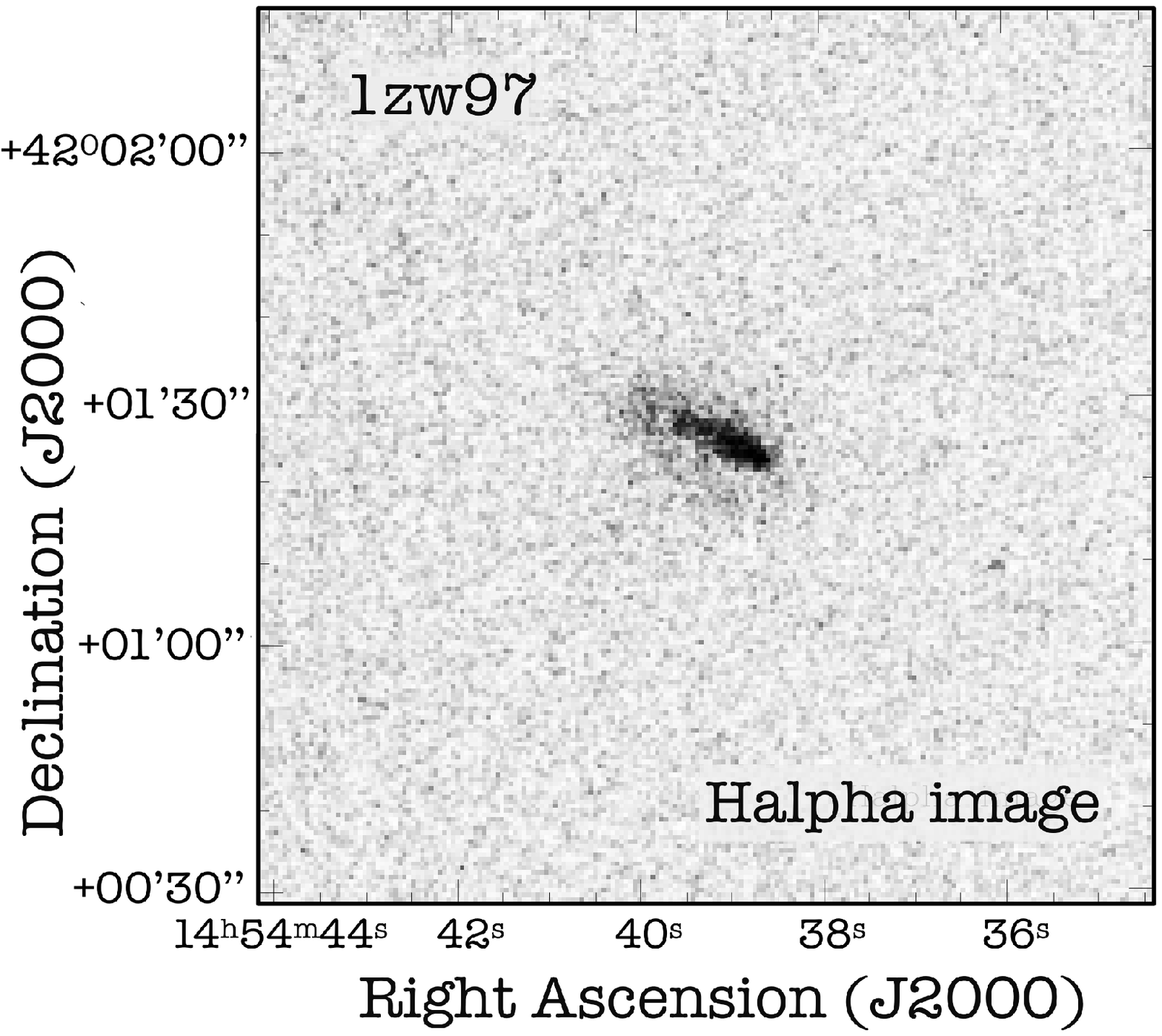}
\caption{Continued ...}
%\label{fig:04}
\end{figure*}

\begin{figure}
\centering
\includegraphics[width=9.0cm, trim=0.7cm 0.9cm 1cm 1.1cm, clip=true]{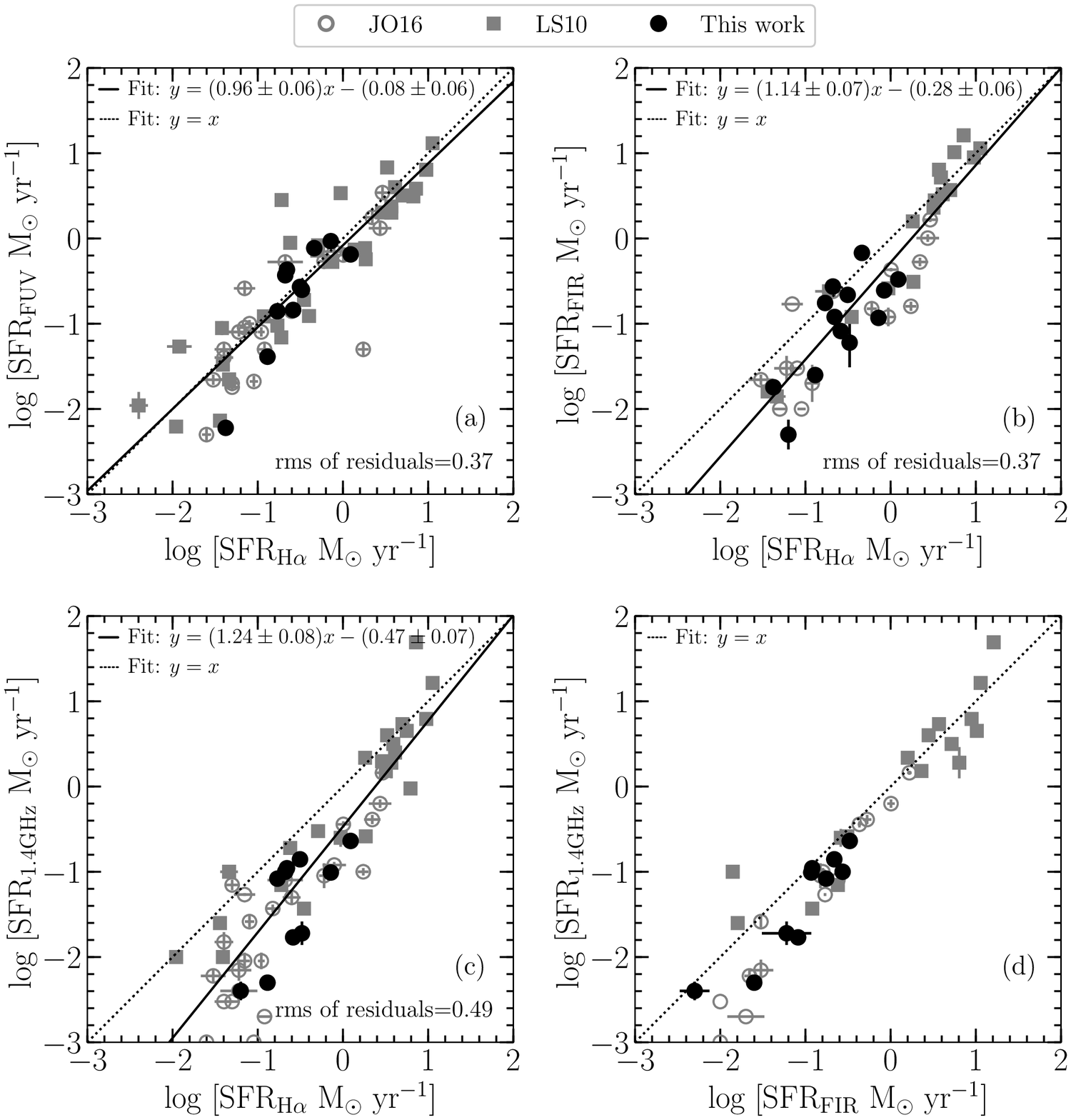}
\caption{Comparison of SFRs derived from various indicators at different wavebands. In each figure, the dotted line represents 1:1 correspondence between two SFRs at different wavelengths and the solid line represents linear fit to the data points.}
\label{fig:02}
\end{figure}

The grey-scale r-band and continuum subtracted H$\alpha$ images of galaxies in our sample are presented in Fig.~\ref{fig:01}. These images show the distribution of star-forming regions and stellar component in galaxies. The H$\alpha$ flux measured from continuum subtracted H$\alpha$ image, corrected for dust extinction using Balmer decrement method and lines contamination are provided in online Table 2. This table also compiles flux measured at FUV, FIR (20, 60 and 100 $\mu$m) and 1.4 GHz radio continuum wavebands. In online Table 3, we provide luminosities and SFRs measured in all wavebands for the studied galaxies. These SFRs were estimated using the same calibration as described in JO16. In order to probe the consistency among various SFR indicators, comparison of H$\alpha$-based SFR with all other SFRs is shown in Fig~\ref{fig:02}. We have included here other 45 WR galaxies from LS10 and JO16 to make a statistically better comparison.        

\begin{figure}
\centering
\includegraphics[width=9.0cm, trim=0.7cm 0.9cm 0.7cm 0.5cm, clip=true]{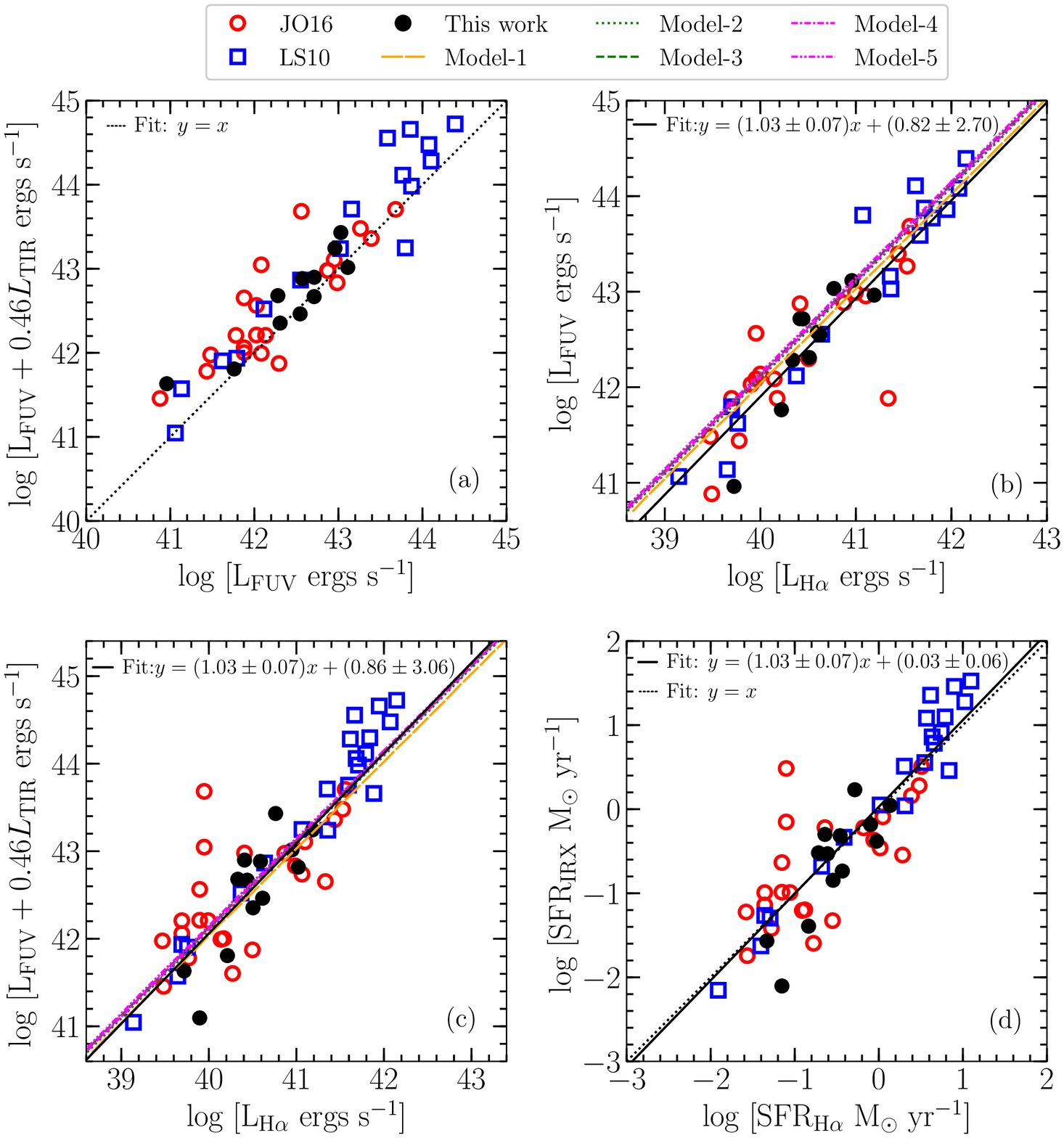}
\caption{(a) Comparison of two FUV luminosities corrected for dust attenuation using IRX and Balmer decrement methods. (b) and (c) show two FUV luminosities corrected for dust using Balmer and IRX method as a function of H$\alpha$ luminosity, compared with various models as named by Model $1-5$ (see text in Sect.~\ref{Sect-3}) based on different SFR prescriptions. (d) Comparison of SFRs derived from the H$\alpha$ and IRX based FUV luminosities.}
\label{fig:03}
\end{figure}

\begin{figure}
\centering
\includegraphics[width=9.0cm, trim=0.7cm 11.2cm 0.7cm 2.4cm, clip=true]{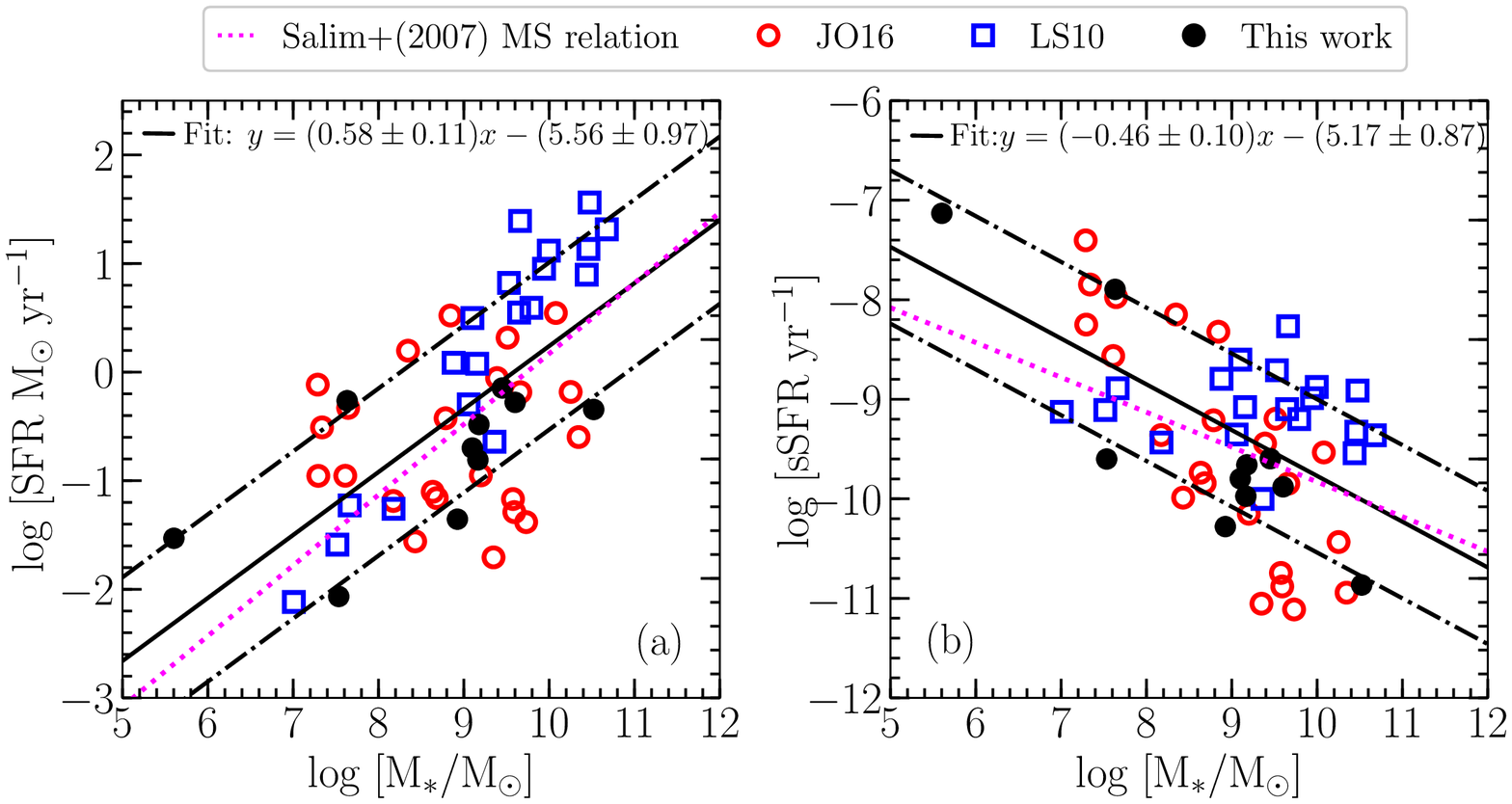}
\caption{(a) The main-sequence (MS; SFR - M$_{*}$) relation for WR galaxies and its comparison with Salim's main-sequence relation. (b) sSFR as a function of stellar-mass of WR galaxies and its comparison with Salim's study.}
\label{fig:04}
\end{figure}

The Balmer decrement based dust extinction correction was done using the SDSS spectroscopic data obtained over a single bright star-forming region within 3" diameter and was assumed uniform throughout the galaxy extent. Such assumption may lead to uncertainty in the dust correction in UV band compared to H$\alpha$ band. Additionally, this method may be sensitive to dust property and its relative distribution around star-forming regions in galaxies. We therefore used an alternative method known as infrared excess (IRX) method which is based on an energetic budget consideration and independent of the dust property \citep{2011ApJ...741..124H}. Fig~\ref{fig:03} (a) shows the comparison between two FUV luminosities corrected for the dust extinction using IRX and Balmer decrement methods. For IRX method, the total IR (TIR) emission is estimated using the FIR fluxes at 25, 60 and 100 $\mu$m using the same prescription given in \citet{2011ApJ...741..124H}. Fig~\ref{fig:03} (b) and (c) shows the comparison of these two FUV luminosities with the H$\alpha$ luminosity, including different models prediction based on various SFR prescriptions. In this figure, different SFR prescriptions are shown as Model $1-5$. Model-1 represents widely used prescription of \citet{1998ARA&A..36..189K} based on old 1990's generation stellar evolution models \citep[see][]{1994ApJ...435...22K,1998ApJ...498..106M}. It assumes a Salpeter IMF with stellar mass limits between $0.1-100$ M$_{\odot}$ and constant star formation history lasting for 100 Myr having solar metallicity. Model-4 and 5 represent the same prescription as adopted in \citet{1998ARA&A..36..189K}, with new stellar population synthesis models of STARBURST99 \citep{1999ApJS..123....3L} and constant star formation history lasting for 100 Myr and 1 Gyr, respectively. Model-2 and 3 are constructed by adopting a more realistic Kroupa IMF \citep{2003ApJ...598.1076K} with mass limits of $0.1-100$ M$_{\odot}$, assuming constant star formation history lasting for 100 Myr and 1 Gyr, respectively. Fig~\ref{fig:03} (d) presents the comparison between IRX and H$\alpha$ based SFRs. In Fig.~\ref{fig:04} (a) and (b), we respectively present the main sequence (MS; SFR-M$_{*}$), and sSFR-M$_{*}$ relations for WR galaxies. In these relations, we used SFRs derived from the IRX method.


%\hspace{-10cm}
\section{Discussion}

In Fig~\ref{fig:01}, it is seen that most of the galaxies (e.g., NGC 1140, NGC 2481, Mrk 94, Mrk 750, IC 745, UGC 6805, NGC 4496 and Izw 97) show one of the following morphological features in their H$\alpha$/r-band images: lopsided stellar and H$\alpha$ light distribution, plumes, tails and arc-like cometary structures, stellar bar and misalignment between ionized H$\alpha$ and stellar disks. Since these features are believed as signs of tidal interaction/merger of galaxies \citep[LS10;][JO16]{2012MNRAS.419.1051L}, we therefore speculate that the studied galaxies are probably tidal interacting/merging candidates. Although the optical interacting/merging counterparts are not seen here, they might be low-luminous dwarfs or \HI clouds. The deep \HI images of studied galaxies are highly desirable to confirm our speculation.

Fig~\ref{fig:02} shows that SFRs from different indicators are well-correlated and improved over previous studies performed in LS10 and JO16. In all the cases, slopes are now close to unity. The strongest correlation is seen between FUV and H$\alpha$-based SFRs with Spearman correlation coefficient of 0.97, while correlation between FIR-H$\alpha$ and radio-H$\alpha$ are identical at 0.96. Assuming that dust extinction plays a greater role in FUV band as compared to H$\alpha$ band, uncertainty in extinction could lead to larger scatter in FUV-H$\alpha$ correlation. Nevertheless, the scatter in FUV-H$\alpha$ and FIR-H$\alpha$ correlations are rather similar, implying that other uncertainty such as variation in IMF and starburst age are main factors for the observed scatter. Moreover, it can be noticed that radio-H$\alpha$ based SFRs show a significant deviation at the low SFR end. This is probably appearing due to the same reason as mentioned in JO16 that WR galaxies are radio deficient due to lack of supernova events in nascent star formation ($\textless$ 10 Myr). 

In Fig~\ref{fig:03} (a), it can be clearly seen that Balmer decrement based FUV luminosity is underestimated compared to IRX based FUV luminosity. In agreement with this finding, it can also be noticed that the correlation between Balmer based FUV and H$\alpha$ luminosity as shown in Fig~\ref{fig:03} (b) is not consistent with different models prediction. However, the correlation between IRX based FUV and H$\alpha$ luminosities are consistent with different models prediction and their predicated SFRs are also well-correlated having slope of $\sim$ 1.03 as shown in Fig~\ref{fig:03} (c) and (d), respectively. This inference implies that the distribution of dust relative to star-forming regions plays an important role and the IRX based SFRs can be used as true SFR estimates. Therefore, the IRX based SFRs were used in our MS and sSFR-M$_{*}$ relations. The consistency between the models and our fitted relation as shown in Fig.~\ref{fig:03} (c) implies that the studied WR galaxies have probably experienced a continuous star formation at least for 1 Gyr over which the recent ($\textless$ 10 Myr) starburst has taken place in the WR phase of galaxies.

The MS relation for normal star-forming galaxies is very tight, although it evolves positively with increasing redshifts \citep{2004MNRAS.351.1151B,2014ApJ...795..104W,2018A&A...609A..82B}. Those systems which show starburst or quenched scenario respectively lie above or below the MS relation, implying that the relation is very sensitive to various physical feedback processes acting in enhancement or quenching of star formation. Fig.~\ref{fig:04} (a) represents the MS relation for nearby WR galaxies which is shown here for the first time in the literature. Similarly, Fig.~\ref{fig:04} (b) represents the relation between sSFR and stellar mass for WR galaxies. In these figures, it can be seen that WR galaxies in our sample including other from LS10 and JO16 show linear relations within 1$\sigma$ scatter of 0.97 dex (as shown by dashed-dot line). This derived relations for WR galaxies are very close to previously known MS relation (with an average scatter of 0.5 dex) drawn by \citet[][;as shown by dotted line]{2007ApJS..173..267S} for nearby normal star-forming galaxies. This analysis implies that WR galaxies in nearby Universe evolve in a similar way as other normal star-forming galaxies evolve. Recently, it is established that galaxy mergers events can change the galaxy's position in either direction of MS relation by only small amount $\sim$ 0.1 dex \citep{2019arXiv190810115P}. Since WR galaxies in the present work are speculated as merging/interacting systems and other from LS10 and JO16 were already confirmed, a slight deviation in slope and a larger scatter in the derived MS relations compared to Salim's relations may therefore be appearing due to merging/interacting nature of the studied systems. However, a large sample of WR galaxies can make this conclusion even more strong.

\section{Conclusions}

The main conclusions of this study are as follows:

\begin{itemize}

\item {The observed H$\alpha$ and r-band morphologies of WR galaxies in our sample revealed the distributions of star-forming regions and stellar component. This morphological study led to speculate that the studied galaxies are potentially tidal interacting and/or merging candidates.}

\item {In the studied WR galaxy sample, the H$\alpha$ based SFRs are found to be well-correlated with SFRs derived using other indicators at FUV, FIR and radio wavebands. Nevertheless, a significant deviation between radio and H$\alpha$ based SFRs at the low SFRs end indicates the deficiency of radio emission in WR galaxies due to lack of supernovae events in nascent starburst.}

\item {Our study suggests that the IRX method gives the best SFR estimate, consistent with the different models prediction. These models suggest that the studied galaxies have probably experienced a continuous star formation at least for 1 Gyr over which the recent ($\textless$ 10 Myr) starburst has taken place in the WR phase of galaxies.}

\item{This study presents the MS relation for nearby WR galaxies for the first time and concludes that WR systems shows a similar linear relation as it is previously known for normal nearby star-forming galaxies in the literature. This finding indicates that WR systems evolve similar to other nearby normal star-forming galaxies. However, the merging/interacting nature of WR galaxies can lead to slight change in slope and a larger scatter in the relation.} 

\end{itemize}

%%%%%%%%%%%%%%%%%%%%%%%%%%%%%%%%%%%%%%%%%%%%%%%%%%%%%%%%%%%%%%%%%%%%%%%%%%%%%%%%%%%%%%%%%%%%%%%%%%%%%%%%%%%%%%%%%%%%%%%%%%%%%%%%%%%%%%%%%%%%%%%%%%%%%%

\section*{Acknowledgements}

We thank the anonymous referee for his/her suggestions that improve the content of manuscript. We thank the staff of ARIES, whose dedicated efforts made these observations possible. 1.3-m DFOT is run by ARIES with support from the Department of Science and Technology, Govt. of India. This research has made use of the NASA/IPAC Extragalactic Database (NED) which is operated by the Jet Propulsion Laboratory, California Institute of Technology, under contract with the National Aeronautics and Space Administration. This research has made use of NASA's Astrophysics Data System.  We acknowledge the use of the $SDSS$ Web Site is http://www.sdss.org/. We gratefully acknowledge NASA's support for construction, operation, and science analysis for the $GALEX$ mission, developed in cooperation with the Centre National d'Etudes Spatiales of France and the Korean Ministry of Science and Technology. The $Infrared$~$Astronomical$~$Satellite$ ($IRAS$) mission was a collaborative effort by the United States (NASA), the Netherlands (NIVR), and the United Kingdom (SERC). VLA (Very Large Array) is run by NRAO (National Radio Astronomy Observatory). The NRAO is a facility of the National Science Foundation operated under cooperative agreement by Associated Universities, Inc. 

\bibliographystyle{mnras} 
\bibliography{references}
\label{lastpage}
\end{document}